\documentclass[prb,twocolumn,showpacs,floatfix,amsmath]{revtex4-1}

\usepackage{color}
\usepackage{dsfont}
\usepackage[normalem]{ulem}
\usepackage{dcolumn}

\usepackage{graphicx}

\newcommand{\unitmatr}{\ensuremath{\mathds{1}}}

\newcommand{\ket}[1]{\ensuremath{\lvert #1 \rangle}}
\newcommand{\bra}[1]{\left\langle#1\right|}

\definecolor{orange}{rgb}{1,0.5,0}
\definecolor{darkgreen}{RGB}{0,100,0}
\definecolor{red}{RGB}{255,0,0}
\definecolor{green}{RGB}{0,255,0}
\definecolor{blue}{RGB}{0,0,255}

\newcolumntype{d}[1]{D{.}{.}{#1}}

\makeatother

\begin{document}

\title{Microscopic theory of the residual surface resistivity of Rashba electrons}
\author{Juba Bouaziz}
\email{j.bouaziz@fz-juelich.de}
\author{Samir Lounis}
\email{s.lounis@fz-juelich.de}
\author{Stefan Bl\"ugel}
\affiliation{$^1$Peter Gr\"unberg Institut and Institute for Advanced Simulation, 
Forschungszentrum  J\"ulich and JARA, 52425 J\"ulich, Germany}
\author{Hiroshi Ishida}
\affiliation{$^2$College of Humanities and Sciences, Nihon University, 
Sakura-josui, Tokyo 156-8550, Japan}
\affiliation{$^3$Center for Materials Research by Information Integration,
National Institute for Materials Science, Tsukuba,
Ibaraki 305-0047, Japan}

\date{\today}

\begin{abstract}
A microscopic expression of  the residual electrical resistivity tensor is derived in linear response theory  
for Rashba electrons scattering at a magnetic impurity with cylindrical or non-cylindrical potential.  The behavior of 
the longitudinal and transversal residual resistivity is obtained analytically and computed for an Fe impurity at the Au(111) surface. We studied 
the evolution  of the resistivity tensor elements as function of the Rashba spin-orbit strength and the magnetization 
direction of the impurity.  We found that the absolute values of longitudinal resistivity reduces with increasing spin-orbit strength of
the substrate and that  the scattering of the conduction electrons at magnetic impurities with magnetic moments pointing in directions
not perpendicular to the surface plane produce a  planar Hall effect and an anisotropic magnetoresistance even if the impurity 
carries no spin-orbit interaction. Functional forms are provided describing the anisotropy of the  planar Hall effect and the anisotropic 
magnetoresistance with respect to the direction of the impurity moment.  
 In the limit of no spin-orbit interaction and a non-magnetic impurity of cylindrical symmetry, the expression of the 
residual resistivity of a two-dimensional electron-gas has the same simplicity and form as for the three-dimensional electron 
gas [J. Friedel, NuovoCimento Suppl.\ {\bf 7}, 287  (1958)] and can also be expressed in terms of scattering phase shifts. 
\end{abstract}

\pacs{72.10.Bg, 72.10.Fk, 72.20.My}
\maketitle

\section{Introduction}

The electron transport in metals and semiconductors is an important
field of research since it crucially influences the efficiency, the power consumption, the size
and the lifetime of electronic components. The use of the spin degree of freedom in
addition to the charge is expected to boost microelectronics by adding new functionalities to existing devices. 
An important building block relating charge to spin currents is the spin-orbit (SO) interaction that appears also 
in terms of the Rashba effect\cite{Rashba:1960,Bychkov:1984} in surface and interface states of heavy metals or in 
semiconductors in contact to those. The Rashba effect arises from the SO coupling in an environment with a lack
of space inversion symmetry such as  interfaces and surfaces where  Bloch-momentum-dependent spin splittings, 
known as  the Rashba spin splitting, are observed in the band structure. The Rashba spin splitting at surfaces 
was first  observed for the two-dimensional (2D) Shockley surface state of the Au(111)
surface.\cite{LaShell, Reinert:01,Blaha:2001,Osterwalder:2004}
It was investigated for a number of clean\cite{Blaha:2001,Bihlmayer:2006,Tamai:2013}
and alloyed surfaces\cite{Bihlmayer:2007, Ast:2007}, adsorbed surface layers\cite{Marchenko:2012,Shikin:2013} and 
surface states of semiconductors in contact to heavy metals.\cite{Gierz:2009} For some of those systems large spin-splittings
had been observed. By this, the Rashba splitting makes possible the  efficient application of the Edelstein 
effect\cite{Edelstein:1990}  mediated creation of large lateral spin-polarization. 
 
The scattering of Rashba electrons at impurities is a source  of magneto-transport properties.
In particular we expect contributions to the  planar Hall effect (PHE), the anomalous Hall effect (AHE) as well as the
anisotropic magnetoresistance (AMR). Both the PHE and AHE are observed as a voltage transverse to the applied 
current\cite{Nagaosa:2010, Seemann:2010} in contrast to the AMR, which is measured 
in the longitudinal geometry. Consequently PHE and AHE are characterized by the transverse resistivity $\rho_{xy}$, 
while the AMR is characterized by the longitudinal one, $\rho_{xx}$. For PHE the in-plane component of the 
magnetization $M_\parallel$ with respect to the interface plane and for  AHE the out-of-plane component $M_\perp$ matter. 
Although the AMR is known since 1856,\cite{Thomson:1857} the PHE was discovered more than a century 
later.\cite{Yau:1971} Only recently it could be established that the PHE can originate from the AMR without contributions from the 
AHE.\cite{Seemann:2011} 

Since the early nineties, with the advent of low-temperature scanning tunnelling microscopy 
(STM) the investigation of the scattering of surface states on impurities  proved very fruitful  for studying the quantum 
behavior of matter. Crommie {\it et al.}\cite{Crommie:1993} reported the direct observation of standing-wave patterns 
due to the scattering of the 2D electron gas off a single Fe impurity on the Au(111) surface. Recently Lounis {\it et al.}\cite{Samir:2012}
refined this picture by including the Rashba effect in the description of the 2D electron gas provided by the Shockley 
surface state of the Au(111) surface and found a magnetic adatom induced skyrmion-like spin texture in the standing wave 
pattern. 

In this paper we study the effect of the SO interaction on the residual resistivity of 2D surface or 
interface Rashba states  induced by a single non-magnetic or magnetic impurity and their contribution to the various 
magneto-transport properties.
The impurity may not necessarily be a single atom. It can be any defect whose potential perturbation is localized.
For example, it can be an adsorbed molecule made out of several atoms.
An objective of this paper is to formulate a residual resistive tensor with longitudinal and 
transverse resistivity contributions as an extension of  a recently derived microscopic linear response expression 
of the surface resistivity for a semi-infinite jellium model\cite{Ishida:1995,Trioni:2002} to the 2D Rashba electron gas.  
The impurity in our model is described by a scattering matrix and we consider in general that the potential perturbation 
induced by the impurity does not need to fulfill any symmetry such as cylindrical
symmetry with respect to the impurity position or a magnetic moment perpendicular to the surface. 
Moreover, the SO interaction can be incorporated at the level of 
the impurity besides the surrounding electron gas (see Ref.\cite{rashba:2007}). In that case the SO coupling contribution to the scattering matrix can be added.
For our application related to a single magnetic atom adsorbed on Au(111), we will, however, neglect this SO term when evaluating 
the residual resistivity tensor. While our focus is on the Rashba electron gas, we note that recently an 
interesting work was performed on the impact of impurities 
on the transport properties of 3D topological insulators surfaces\cite{Parijat2015}.

We found that 
in absence of SO interaction and of the magnetism at the impurity, the diagonal contributions of the resistivity tensor
induced by a single adatom 
can be expressed in terms of scattering phase shifts just alike the  well-known expression by Friedel\cite{Friedel:1958},
\begin{equation}
\label{eq:RR_3D-electron-gas}
\rho = A_\mathrm{F}  \sum_{\ell \geq 0} (\ell+1) \sin^2
[\delta_{\ell+1}(\varepsilon_\mathrm{F}) - \delta_{\ell}(\varepsilon_\mathrm{F})]\, ,
\end{equation}
where the  residual resistivity  $\rho$, at $T = 0$~K induced 
by a non-magnetic impurity with a spherical potential in a 3D 
degenerate free-electron gas is related to the momentum-transfer cross section of electrons at the Fermi surface 
given by a sum over the orbital momentum quantum 
numbers $\ell$. Here $A_\mathrm{F}$ is a constant prefactor proportional to the inverse of the wave vector $k_\mathrm{F}$ 
at the Fermi energy $\varepsilon_\mathrm{F}$ of the host conduction electrons described by
the free 3D electron gas. $\delta_l(\varepsilon_\mathrm{F})$ is the phase shift  between the scattered wave
function at the impurity and the unscattered host wave function.

 As an application we have considered an Fe 
impurity on the  Au(111) surface, both non-magnetic and magnetic with perpendicular and arbitrary direction of the 
impurity moment with respect to the surface plane. 
If the impurity moment has a component parallel to the surface plane the scattering matrix as well as the transverse 
components of the resistivity tensor are non-zero even when the impurity potential has cylindrical symmetry
and a PHE is found. We follow the longitudinal resistivity and the AMR as function of the spin-orbit strength and the 
transverse resistivity and the PHE as function of the orientation  of the magnetic impurity moment with respect to the
surface plane. The AHE is absent since we consider a non-spinpolarized 2D Rashba electron gas 
\cite{Nagaosa:2010} and is not further considered in this paper. 
 Also, we provide a phenomenological functional form for the different components of the residual resistivity as 
function of the orientation of the magnetic moment. 

The paper is organized as follows: In Sec.~\ref{sec:Theory} we briefly introduce the Rashba model, basically to 
define all quantities. Then, the ingoing and outgoing scattering states of the Rashba Hamiltonian are introduced.
We express the resistivity tensor components in terms of  scattering coefficients, which we relate to the scattering 
matrix via the Lippmann-Schwinger equation. We introduce expressions for cylindrical and non-cylindrical impurity 
potentials of magnetic impurity moments normal to the surface and arbitrary orientation of the magnetic impurity moment.
The result of the resistivity tensor {\it i.e.}\ the application of the derived expression for an Fe impurity on Au(111) is reported and discussed in Sec. III.
 A summary is provided in Sec.~\ref{sec:Summary}.

\section{Theory}
\label{sec:Theory}
\subsection{Rashba Model}
The SO interaction leads in a structure-asymmetric environment
 such as a surface or interface  to a spin-splitting of the otherwise
 two-fold degenerate eigenstates of a two-dimensional electron gas. 
 The model of Bychkov and Rashba\cite{Rashba:1960,Bychkov:1984} describes 
this splitting by adding to the kinetic energy of the free electrons the  
so-called Rashba Hamiltonian
\begin{equation}
\label{eq:Rashba_Ham}
\mathbf{H} = \frac{ p^2_{x} + p^2_{y} }{2m^*}\otimes \unitmatr_2 
                  - \frac{\alpha_\mathrm{so}}{\hbar}(\boldsymbol{\sigma}_{x} {p_{y}} - \boldsymbol{\sigma}_{y} {p_{x}})\, ,
\end{equation}
where ${p_{\gamma}}$, $\gamma \in \{x,y\}$, are the components of the momentum
operator $\vec{p}$ in a cartesian coordinate system with $x,y$ coordinates in the surface plane whose surface normal 
points along $\hat e_z$.  $m^*$ is the effective mass of the electron. 
$\boldsymbol{\sigma}_\gamma$ are the Pauli matrices and $\unitmatr_2$ is the unit matrix in spin-space
with a global spin frame of reference  where the spin z-direction is aligned parallel to $\hat e_z$.
$\alpha_\mathrm{so}$ is the Rashba parameter,  a measure of the strength of the SO
interaction and the parameter that controls the degree of Rashba spin splitting.

The eigenstates corresponding to this Hamiltonian are written as a product of a plane wave in space and a two-component
spinor
\begin{equation}
{\psi_{\vec{k},\pm}}(\vec{r}) = \frac{1}{\sqrt{2}} e^{i\vec{k}\cdot\vec{r}}
 \left(
\begin{array}{c}
1 \\
\pm ie^{i\phi_{\vec{k}}} \\
\end{array}
\right)\>\textrm{with}\> \phi_{\vec{k}}=\arctan\left(\frac{k_y}{k_x}\right),
\label{Rashba_eigen}
\end{equation}
{\it i.e.}\ they can be considered as a superposition of spin-up and down-states  when measured with respect to the
surface normal. The orientation of the local spin-quantization axis, given by the expectation value 
$\hat{n}_{\pm}(\vec k) = \langle  \psi_{\pm \vec{ k}}  \vert \vec{\mbox{\boldmath $\sigma$}} \vert 
                                                                      \psi_{\pm \vec{k}} \rangle 
= \pm (-\sin\phi_{\vec{k}},  \cos\phi_{\vec{k}}, 0) =\pm {\hat e}_\phi$
lies in the surface plane and is perpendicular to the wave vector 
$\vec{k}=k(\cos\phi_{\vec{k}} , \sin\phi_{\vec{k}}, 0)=k\,{\hat e}_k$. We find that the quantization axis is independent of the magnitude 
$k$ and depends only on the direction ${\hat e}_k$ of the wave vector $\vec k$. With respect to this quantization axis
that is parallel to ${\hat e}_\phi $ in a cylindrical coordinate system, $\psi_\alpha$ are spin pure eigenstates and we can
associate $\psi_\alpha$ for $\alpha=+ (-)$ as spin-up (-down) state.  

The energy dispersion is characterized by the $k$-linear splitting of the free-electron parabolic band dispersion as denoted:
\begin{equation}
\varepsilon_{\pm}(k) = \frac{\hbar^2 k^2}{2m^*}
 \pm \alpha_\mathrm{so} k = \frac{\hbar^2 }{2 m^*} \left[(k \pm  k_\mathrm{so})^2 - k_\mathrm{so}^2\right].
\label{energy_dispersion}
\end{equation}

Due to the presence of the $z$-inversion broken symmetry and 
the SO interaction, the origins of the spin-up and -down parabola is shifted
by the  Rashba or the spin-orbit wave vector, respectively,  $k_\mathrm{so} = {m^*\alpha_\mathrm{so}}/{\hbar^2}$
and the dispersion relation describes two bands. 

\subsection{Scattering States}
\label{sec:IIB}
In order to describe the scattering problem of the Rashba electrons at a single impurity,
it is convenient to exploit the cylindrical symmetry of the Rashba electron gas,
to introduce the cylindrical coordinates $(r,\phi)$ with radius $r$ and azimuth 
$\phi$ between the vector $\vec{r}$ and the $x$ axis, and to place the impurity at the origin of the coordinate
system. For this purpose it is more convenient to express the eigenfunctions of the Rashba
Hamiltonian~\eqref{eq:Rashba_Ham} in terms of the cylindrical Bessel functions rather than plane waves as
\begin{equation}
\label{eq:regsol}
\psi_{\varepsilon m \pm}(\vec{r}) =  \frac{e^{-\frac{2m+1}{4}\pi i}}{\sqrt{2}} 
\left(
\begin{array}{c}
J_{m}(k_\pm r) e^{im\phi}\\
\mp J_{m+1}(k_\pm r)  e^{i(m+1)\phi}\\
\end{array}
\right),
\end{equation}
where the state is labeled by the energy $\varepsilon$, band index $\alpha=\pm$,
and the orbital angular quantum number $m$, $m\in \mathds{Z}$,  rather than the wave vector
$\vec k$ or $(k, \phi_{\vec{k}})$.
 The wave number for band $\alpha$ is defined by
\begin{eqnarray}
 \left\{ \begin{array}{ll}
 k_\pm = k_{\mathrm{M}} \mp k_\mathrm{so},
 \\ \nonumber
 \mathrm{with}\  k_{\mathrm{M}} = \left[\frac{2m^*\varepsilon}{\hbar^2}+k_\mathrm{so}^2\right]^{1/2},
 \end{array}\right.\label{k+-}\hspace{5ex}
 \end{eqnarray}
so that $k_{+}-k_{-}=-2k_\mathrm{so}$ holds irrespective of the value of  $\varepsilon (>0)$.
We note that
 $\psi_{\varepsilon m \alpha}(\vec{r})$ is an eigenvector for the $z$
component of the total angular momentum operator $\mathbf{j}_{z}  = 
\mathbf{l}_{z}+\frac{\hbar}{2} \boldsymbol{\sigma}_{z} $ with an
eigenvalue $j_\mathrm{z}  = \hbar (m+\frac{1}{2})$ and $\mathbf{l}_{z}$ is the orbital angular momentum operator. 
${\psi_{\varepsilon 
m \alpha}}(\vec{r})$ can be decomposed into an incident and an outgoing wave:
$\psi_{\varepsilon m \alpha}(\vec{r}) = \psi_{\varepsilon m \alpha}^\mathrm{in}(\vec{r})
+ e^{-(m+\frac{1}{2}) \pi i} \psi_{\varepsilon m \alpha}^\mathrm{out}(\vec{r})$.
For each band component, the incoming and outgoing
wave functions are respectively cylindrical Hankel functions of second kind (see Ref.\cite{Walls2006} 
for a similar derivation),
\begin{equation}
{\psi_{\varepsilon m \pm}^\mathrm{in}}(\vec{r}) =
 \frac{e^{-\frac{2m+1}{4} \pi i}}{2\sqrt{2}} 
\left(
\begin{array}{c}
\hspace{-1ex} H_{m}^{(2)}(k_{\pm}r) e^{im\phi}\\
\mp H_{m+1}^{(2)}(k_{\pm}r)  e^{i(m+1)\phi}\\
\end{array}
\right), \label{eq:inwv}
\end{equation}
and first kind,
\begin{equation}
{\psi_{\varepsilon m \pm}^\mathrm{out}}(\vec{r}) =
\frac{e^{\frac{2m+1}{4} \pi i}}{2\sqrt{2}} 
\left(
\begin{array}{c}
H_{m}^{(1)}(k_{\pm}r) e^{im\phi}\\
\mp H_{m+1}^{(1)}(k_{\pm}r)  e^{i(m+1)\phi}\\
\end{array}
\right)\, .\label{eq:outwv}
\end{equation}
Their phase factors are chosen such that at
large distances ($r \rightarrow \infty$) we can express them as:
\begin{equation}
\psi_{\varepsilon m \pm}^\mathrm{in}(\vec{r})  =  \frac{1}{\sqrt{4 \pi k_{\pm} r}} e^{-i k_{\pm} r} e^{im\phi}
\left(\begin{array}{c}
1\\
\mp i e^{i\phi}
\end{array}\right),
 \end{equation}
\begin{equation}
\psi_{\varepsilon m \pm}^\mathrm{out}(\vec{r})  =  \frac{1}{\sqrt{4 \pi k_{\pm} r}} e^{i k_{\pm} r} e^{im\phi}
\left(\begin{array}{c}
1\\
\pm i e^{i\phi}
\end{array}\right),
\end{equation}
with $\psi_{\varepsilon m \pm}^\mathrm{in}$ and $\psi_{\varepsilon m \pm}^\mathrm{out}$
describing 2D cylindrical waves incoming toward and outgoing from the origin of the coordinate
system,  respectively. They are related by
\begin{equation}
\hat{T} \psi_{\varepsilon m \pm}^\mathrm{in}(\vec{r})
=\mp i\ \psi_{\varepsilon,\ -(m+1),\ \pm}^\mathrm{out}(\vec{r}), \nonumber
\end{equation}
where $\hat{T}$ denotes the time reversal operator. 

Now, we introduce a localized impurity
for convenience placed at the origin of the cylindrical coordinate 
system of the 2D electron gas. We describe the elastic scattering of the wave function
$\psi_{\varepsilon m \pm }$ from the impurity with the Lippmann-Schwinger
 equation involving real and spin space coordinates:
\begin{eqnarray}
\varphi_{\varepsilon m \alpha}(\vec{r})&=& 
\psi_{\varepsilon m \alpha}(\vec{r}) \nonumber \\
+\int &d\vec{r}^{\,\prime}&
  d\vec{r}^{\,\prime \prime} \hspace{1mm} \mathbf{G}_{0}(\vec{r}, \vec{r}^{\,\prime}, \varepsilon)
   \mathbf{t}(\vec{r}^{\,\prime}, \vec{r}^{\,\prime \prime})
 \psi_{\varepsilon m \alpha}(\vec{r}^{\,\prime \prime}), \label{eq:lippschwi_tmat_r}
\end{eqnarray}
where $\mathbf{G}_{0}(\vec{r}, \vec{r}^{\,\prime}, \varepsilon)$ is the Green function
of the Rashba electron gas and $\mathbf{t}(\vec{r}, \vec{r}^{\,\prime})$
 is the transition matrix ($t$-matrix), related to the impurity potential $\mathbf{v}(\vec{r})$ via the Dyson
 equation: $\mathbf{t}(\vec{r}, \vec{r}^{\,\prime}) = \mathbf{v}(\vec{r}) \ \delta(\vec{r}-\vec{r}^{\,\prime})
+ \int d\vec{r}^{\,\prime \prime}\ \mathbf{v}(\vec{r})
\ \mathbf{G}_{0}(\vec{r}, \vec{r}^{\,\prime \prime}, \varepsilon)
\ \mathbf{t}(\vec{r}^{\,\prime \prime}, \vec{r}^{\,\prime})$.
In the asymptotic region where  the impurity potential $\mathbf{v}(\vec{r})$ vanishes,
 Eq.~\eqref{eq:lippschwi_tmat_r} can be written in a simpler form by using
scattering coefficients $C(m \alpha, m^\prime \alpha^\prime)$. 
Then, the wave function of an incident electron with quantum state $(\varepsilon, m, \alpha)$ scattering
elastically from a non-cylindrical  impurity potential placed at the origin is expressed as
\begin{equation}
\varphi_{\varepsilon m \alpha}(\vec{r}) = 
\psi_{\varepsilon m \alpha}^\mathrm{in}(\vec{r}) + \sum_{m^\prime,\alpha^\prime}
\sqrt{\frac{k_{\alpha^{\prime}}}{k_{\alpha}}}\ C(m \alpha, m^\prime \alpha^\prime) \psi_{\varepsilon m^\prime \alpha^\prime}
^\mathrm{out}(\vec{r}).
\label{eq:der1}
\end{equation}
where the factor $\sqrt{k_{\alpha^\prime}/k_\alpha}$ accounts for the fact that the incoming and
outgoing waves, $\psi_{\varepsilon m \alpha}^\mathrm{in}$ and
$\psi_{\varepsilon m \alpha}^\mathrm{out}$,
carry electron current
$k_M/k_\alpha$
rather than unity due to the relativistic correction of the velocity
operator which will be discussed below.
Here, the scattering coefficients fulfill the unitary condition,
\begin{equation}
\sum_{m_{1} \alpha_{1}} C(m\alpha,m_1\alpha_1)\ C^*(m^\prime \alpha^\prime,m_1\alpha_1)=\delta_{m m^\prime}
\delta_{\alpha\alpha^\prime},
\end{equation}
Specifically, the diagonal elements
of the above equation with $m=m^\prime$ and $\alpha=\alpha^\prime$,
\begin{equation}
 \sum_{m_1\alpha_1}| C(m \alpha, m_1\alpha_1) |^2=1,
\end{equation}
ensure a current conservation.

For $\alpha\ne \alpha^\prime$ the coefficients give weight to the inter-band transition during the scattering.
For $m\ne m^\prime$, the direction of $\vec{k}$ and thus the total angular momentum component of the Rashba electrons 
changes during the scattering process,
and the scattering coefficients refer to the amplitude of the intra-band scattering.
When $\mathbf{v}(\vec{r})$ has a cylindrical symmetry, {\it i.e.}\ $\mathbf{v}(\vec{r})=\mathbf{v}({r})$,
the orbital quantum number $m$ is conserved and $C(m \alpha, m^\prime \alpha^\prime)$ simplifies
to $C(m \alpha, m^\prime \alpha^\prime)\delta_{m,m^\prime}$. 
The scattered wave function will be a linear combination of the spin-splitted
eigenstates denoted by the $+$ and $-$ bands. This mixing is due to the spin-flip inter-band
transitions whose origin is  the off-diagonal part of the Rashba Green $\mathbf{G}_{0}$ function coming from
the spin-orbit interaction. 

\subsection{Connection to the scattering matrix}
\label{sec:tmat_connec}
We present the relation between the scattering coefficients $C(m\alpha, m \alpha^\prime)$ 
and the $t$-matrix $\mathbf{t}(\vec{r}, \vec{r}^{\,\prime})$ elements in the orbital momentum  representation.
For this purpose, it is convenient to express the Rashba Green function in terms of solutions of the Rashba
Hamiltonian~\eqref{eq:Rashba_Ham} in the cylindrical coordinate system presented in Sec.\ \ref{sec:IIB}.
In order to derive the Green function, we fix $\vec{r}^{\,\prime}$ and consider
 $\mathbf{G}_{0}(\vec{r}, \vec{r}^{\,\prime}, \varepsilon)$ to be a function of  $\vec{r}$.
Then, $\mathbf{G}_{0}(\vec{r}, \vec{r}^{\,\prime}, \varepsilon)$ are found to be  a linear combination of
the solutions given by Eq.~\eqref{eq:regsol} and the out-going solutions (so-called irregular solutions) given by 
Eq.~\eqref{eq:outwv}.
Furthermore, by taking account of the cusp condition of the Green function at $r=r^\prime$, it is easy to
derive,
\begin{eqnarray}
 &\mathbf{G}_{0}&(\vec{r},\vec{r}^{\,\prime}, \varepsilon) =  \frac{2 }{i(k_{+} + k_{-})} \label{eq:G0} \\
       &\times& 
 \left\{ \begin{array}{ll}
 \sum_{m \alpha} k_{\alpha} \ket{\psi^\mathrm{out}_{\varepsilon m \alpha}} \bra{\psi_{\varepsilon m \alpha}}e^{-i(m + \frac{1}{2})\pi}
 & \textrm{for $r > r^\prime$}, \\ \nonumber
 \sum_{m \alpha} k_{\alpha} \ket{\psi_{\varepsilon m \alpha}}\bra{\psi^\mathrm{out}_{\varepsilon m \alpha}}e^{i(m + \frac{1}{2})\pi}
 & \textrm{for $r < r^\prime$}.
 \end{array}\right.\hspace{5ex}
 \end{eqnarray}
Using Eqs.~\eqref{eq:G0} and \eqref{eq:lippschwi_tmat_r},  one arrives at this general
expression for the scattering coefficients:
\begin{equation}
\begin{split}
C(m\alpha, m^\prime \alpha^\prime)&= 
\bigg{[}\delta_{mm^\prime} \delta_{\alpha\alpha^\prime} + \frac{2k_{\alpha^\prime}} {i (k_{+} + k_{-})} \\
&\times \bra{\psi_{\varepsilon m^\prime \alpha^\prime}} \mathbf{t}\ket{\psi_{\varepsilon m \alpha}}\bigg{]}\ 
\sqrt{\frac{k_{\alpha}}{k_{\alpha^{\prime}}}}
 e^{-i(m^\prime +\frac{1}{2}) \pi} .
\end{split}
\label{eq:lipp_r}
\end{equation}	

\subsection{Residual resistivity tensor}
Using the Kubo linear response formalism \cite{Ishida:1995} we can show 
that the components of the resistivity tensor $\rho_{\gamma \gamma^\prime}$ measuring the potential drop in direction 
$\gamma$ after applying an electric field in direction $\gamma^\prime$ in the DC limit $\omega$ $\rightarrow 0$
are given  in terms of the scattering solution $\varphi_{\varepsilon m \alpha}$ by 
\begin{equation}
\begin{split}
\rho_{\gamma \gamma^\prime} & = \lim_{\omega \rightarrow 0} 
\frac{\pi \omega}{S n_\mathrm{e}^2 e^2} 
\sum_{i,j} \delta (\varepsilon_{j}-\varepsilon_{i}-\hbar\omega) 
(f_{i}-f_{j})\\
&\hspace{20mm} 
\times \bra{\varphi_i} m^{*}v_{\gamma} \ket{\varphi_j} 
\bra{\varphi_j} m^{*}v_{\gamma^\prime} \ket{\varphi_i},
\end{split}
\label{eq:kubo_form_r}
\end{equation}
where $\gamma,\gamma^\prime \in \{{x}, {y}\}$,
the indices $i,j$ stand each for  $(\varepsilon, m, \alpha)$, $n_\mathrm{e}$ is the
surface electronic density, $S$ denotes the area of the surface, $e$ is the electron
charge, and $f_{i}=\theta(\varepsilon_F-\varepsilon_{i})$ is the occupation number for
the energy level $\varepsilon_{i}$ at $T = 0 $ K. 
Here, $n_e$ is related to the Fermi wave numbers of the two bands,
$k_{F+}$ and $k_{F-}$, by
\begin{equation}
n_e = \frac{1}{4\pi}\left(k_{F+}^2 + k_{F-}^2\right),
\label{Density}
\end{equation}
and  the factor $1/S$ in Eq.~\eqref{eq:kubo_form_r} may be regarded as representing the impurity
number density, $n_i$, if $n_i$ is low enough. Also, one needs in Eq.~\eqref{eq:kubo_form_r} the relativistic velocity operator, {\it i.e.},
\begin{eqnarray}
 \left\{ \begin{array}{ll}
  v_{x} = -i\frac{\hbar}{m^*}\ \frac{\partial}{\partial x} +\frac{\hbar}{m^*}k_{\mathrm{so}} \sigma_{y},
 \\ \nonumber
  v_{y} = -i\frac{\hbar}{m^*}\ \frac{\partial}{\partial y} - \frac{\hbar}{m^*}k_{\mathrm{so}} \sigma_{x}.
 \end{array}\right.\hspace{5ex}
\label{relativistic_momentum}
 \end{eqnarray}
In addition to the prefactor $\omega$, the summation over states $i$ and $j$ in Eq.~\eqref{eq:kubo_form_r}
gives rise to another factor $\omega$, since  $\varepsilon_i$ must satisfy the condition,
$\varepsilon_{F}-\hbar\omega\leq \varepsilon_{i}\leq \varepsilon_{F}$. In spite of this,
the right-hand side of Eq.~\eqref{eq:kubo_form_r} takes a finite limiting value
 in the limit of $\omega\to0$, since the matrix elements
$\bra{\varphi_i} m^{*}v_{\gamma} \ket{\varphi_j}$ evaluated for the asymptotic scattering region
($r\rightarrow \infty$) diverge as $1/\omega$ in the limit of $\omega\to0$, as will be
demonstrated in Appendix~\ref{sec:Appendix_A}. 

The resistivity tensor is related to the energy dissipation $P$ in the system per unit time by
\begin{equation}
P=\sum_{\gamma, \gamma^\prime} \rho_{\gamma \gamma^\prime} J^{*}_\gamma J_{\gamma^\prime},
\label{dissipation}
\end{equation}
where $J_\gamma$ denotes the 2D current density in the $\gamma$ direction.
By inserting the matrix elements of the momentum operators in the limit of $\omega\to 0$
given in Appendix A into Eq.~\eqref{eq:kubo_form_r}, one can derive the most general expression
for the diagonal components of the resistivity tensor,
\begin{eqnarray}
\nonumber
\rho_{\gamma \gamma}  =  \frac{\hbar k^{2}_{\mathrm{FM}} }{8 \pi S
 n_\mathrm{e}^2 e^2} & &\!\!\!\!\!\!\! \sum_{m\alpha, m^\prime \alpha^\prime} 
\Big\vert(\delta_{m\prime,m+1}\pm\delta_{m\prime,m-1})\delta_{\alpha \alpha^\prime} \\  \nonumber
&  + &\! \sum_
{l\alpha^{\prime\prime}} C(m \alpha, l \alpha^{\prime\prime}) C^{*}(m^\prime \alpha^\prime, l+1 \alpha^{\prime\prime}) \\ 
& \pm &\!\sum_{l\alpha^{\prime\prime}} C(m \alpha, 
l \alpha^{\prime\prime}) C^{*}(m^\prime \alpha^\prime, l-1 \alpha^{\prime\prime})  \Big\vert^{2}, 
\label{eq:der7}
\end{eqnarray}
where $k_{\mathrm{FM}} = k_{\mathrm{M}}(\varepsilon_{\mathrm{F}})$ 
and the plus and minus signs correspond to $\rho_{xx}$ and $\rho_{yy}$, respectively.
In deriving Eq.~\eqref{eq:der7}
we replaced  in~\eqref{eq:kubo_form_r} the  sum $\sum_{i}$ over discrete states by 
$\frac{1}{2\pi} \sum_{m=-\infty}^{+\infty} \sum_{\alpha=\pm} \int_0^{+\infty}  k_{\alpha} dk_{\alpha}$. 
The scattering coefficients $C(m\alpha, m^{\prime} \alpha^\prime)$ are implicitly energy dependent. 
After integrating over $k_\alpha$ these coefficients will be taken at the Fermi level, $\varepsilon_\mathrm{F}$.
In the case of the absence of the impurity, the resistivity vanishes since the scattering coefficients are
given in this case simply by $C(m\alpha,m^\prime\alpha^\prime)=\delta_{mm^\prime}\delta_{\alpha\alpha^\prime}
e^{-i(m+1/2)\pi}$.

 If we assume that the impurity has no on-site SO coupling and also that it has a magnetic moment
perpendicular to the surface, then the $t$-matrix is diagonal in spin space
\begin{equation}
\mathbf {t}=
\left(
\begin{array}{rcl}
\-t_{\uparrow \uparrow} &\-0 \\ 
\-0&\-t_{\downarrow \downarrow}\\
\end{array}
\right),\label{eq:t_diagonal}
\end{equation} 
{\it i.e.}\ spin-up and spin-down electrons scatter differently at the impurity. However, the scattering is not 
spin conserving, because the spin of the Rashba electrons lies in the plane and the 
$t$-matrix is not diagonal anymore in that 
spin frame of reference.
Furthermore, if the impurity potential is cylindrical, the orbital  momentum representation of
 $t_{\sigma \sigma}$ with $\sigma \in \{\uparrow, \downarrow\}$ reads
\begin{equation}
t_{\sigma \sigma}(\vec{r}, \vec{r}^{\,\prime}) = \frac{1}{2 \pi} \sum_{m} 
e^{im \phi} t_{\sigma \sigma,m}(r,r^\prime) e^{-im \phi^\prime}.\label{eq:t_cyl}
\end{equation}
In this case, as seen from Eq.~\eqref{eq:lipp_r}, the scattering coefficient
$C(m \alpha, m^\prime \alpha^\prime)$ becomes diagonal with respect to $m$ and $m^\prime$,
and the expression for the resistivity, Eq.~\eqref{eq:der7}, is further simplified as
\begin{eqnarray}
\nonumber
\rho_{\gamma \gamma} & =&  \frac{\hbar k^{2}_{\mathrm{FM}}}{8 \pi S  n_\mathrm{e}^2 e^2}
\sum_{m\alpha, m\prime=m\pm1\ \alpha^\prime}\\ 
&\times& \Big\vert \delta_{\alpha \alpha^\prime} + \sum_{\alpha^{\prime\prime}}
C(m \alpha, m \alpha^{\prime\prime}) C^{*}(m^\prime \alpha^\prime, m^\prime \alpha^{\prime\prime})
\Big\vert^{2}. \label{eq:der8}
\end{eqnarray}
Obviously, the two diagonal components, $\rho_{xx}$ and $\rho_{yy}$, are identical in this case.

\subsection{In the limit of $k_\mathrm{so}=0$}

Here, we consider the limit of  $k_\mathrm{so} \rightarrow 0$ to derive the expression of the
residual resistivity induced by a localized impurity for the 2D free-electron gas without the Rashba SO term.
This may be useful since the derivation of this quantity has not appeared in the literature to our
knowledge. For this purpose, it is better to choose energy $\epsilon$, orbital angular momentum
$m$, and spin index $\sigma$ as the quantum numbers for the description of scattering states,
where the spin quantization axis is chosen as the $z$ axis as in previous sections.
Namely, instead of  Eqs.~\eqref{eq:inwv} and \eqref{eq:outwv}, we employ
\begin{equation}
\tilde{\psi}_{\varepsilon m \sigma}^\mathrm{in}(\vec{r}) =\frac{1}{2}
e^{-\frac{2m+1}{4} \pi i} H_{m}^{(2)}(k r) e^{im\phi} |\sigma\rangle,
\label{eq:inwv2}
\end{equation}
and
\begin{equation}
\tilde{\psi}_{\varepsilon m \sigma}^\mathrm{out}(\vec{r}) =\frac{1}{2}
e^{\frac{2m+1}{4} \pi i} H_{m}^{(1)}(k r) e^{im\phi} |\sigma\rangle,
\label{eq:outwv2}
\end{equation}
as the incident and scattered electron wave functions, where $k=\sqrt{2m^* \varepsilon}/\hbar$,
$|\uparrow\rangle=(1,0)$, and $|\downarrow\rangle=(0,1)$.

Then, the wave function of an incident electron with quantum state $(\varepsilon, m, \sigma)$ scattering
elastically from a non-cylindrical  impurity potential placed at the origin is expressed as
\begin{equation}
\tilde{\varphi}_{\varepsilon m \sigma}(\vec{r}) = 
\tilde{\psi}_{\varepsilon m \sigma}^\mathrm{in}(\vec{r}) + \sum_{m^\prime,\sigma^\prime}
 \tilde{C}(m \sigma, m^\prime \sigma^\prime) \tilde{\psi}_{\varepsilon m^\prime \sigma^\prime}
^\mathrm{out}(\vec{r}).
\label{eq:der9}
\end{equation}

By following the same procedure, we can easily show that the resistivity tensor for the present
case with $\alpha_\mathrm{so}=0$ is given exactly in the same form as Eq.~\eqref{eq:der7}, except that
$k_{F\alpha}$ is replaced by $k_\mathrm{F}=\sqrt{2m^*\varepsilon_\mathrm{F}}/\hbar$,
the scattering coefficients  of the type $C(m \alpha, m^\prime \alpha^\prime)$ are replaced by
$\tilde{C}(m \sigma, m^\prime \sigma^\prime)$, and further the summation over band indices
is replaced by the one over spin indices.

Furthermore, if the $t$-matrix is diagonal with respect to electron spin and the impurity potential has
cylindrical symmetry, we can derive a more simplified expression corresponding to Eq.~\eqref{eq:der8},
\begin{equation}
\rho_{\gamma \gamma}  =  \frac{\hbar k_\mathrm{F}^2}{8 \pi S  n_\mathrm{e}^2 e^2}
\sum_{m, m\prime=m\pm1, \sigma} 
\Big\vert 1 + \tilde{C}(m \sigma, m \sigma) \tilde{C}^{*}(m^\prime \sigma, m^\prime \sigma)
\Big\vert^{2}, 
\label{eq:der10}
\end{equation}   
where it should be noted that in contrast to the Rashba electrons with a finite
$\alpha_\mathrm{so}$, the spin-flip scattering does not occur in the present case.
The scattering coefficient in the above equation can be expressed by using the phase shift as
\begin{equation}
\tilde{C}(m \sigma, m \sigma)=e^{-(m+\frac{1}{2})\pi i + 2\delta_m(\varepsilon,\sigma) i}.
\end{equation}
By inserting this expression into Eq.~\eqref{eq:der10}, we obtain finally
\begin{equation}
\rho_{\gamma \gamma} = \frac{ 2\hbar}{S n_\mathrm{e} e^{2}} 
\sum_{\sigma=\uparrow \downarrow} \sum_{m=-\infty}^{+\infty} \sin^2\left[\delta_{m+1}
(\varepsilon_\mathrm{F},\sigma) - \delta_{m}(\varepsilon_\mathrm{F},\sigma)\right],
\label{eq:res_nocsoc_r}
\end{equation}
where we used the relation $k_\mathrm{F}^2=2\pi n_e$.
This is a modification of Friedel's result~\cite{Friedel:1958} for the 
residual resistivity of a single impurity in a 3D electron gas
(see Eq.~\eqref{eq:RR_3D-electron-gas}) to the case of an impurity in a 2D electron
gas without the Rashba-type SO term. The only difference is the scattering phase space
of momentum transfer in the field direction, which is larger in the  3D case than in 2D and this is
taken care of in Eq.~\eqref{eq:RR_3D-electron-gas} by the multiplicity  $\ell+1$ of each 
angular momentum component.

\subsection{$s$-wave scatterer}
\label{subsec_swave}

In this section we will consider the scattering of  Rashba electrons by an impurity
whose spatial extent is much smaller than the Fermi wave length. For such a scatterer,
one may be able to employ the $\delta$-function approximation for the $t$-matrix,
\begin{equation}
t_{\sigma \sigma^\prime}(\vec{r}, \vec{r}^{\,\prime}) = \delta(\vec{r})\delta(\vec{r}^{\,\prime})
\ t_{\sigma \sigma^\prime}(\varepsilon).
\label{eq:t_delta}
\end{equation}
It should be noted that within this $s$-wave approximation, only $\psi_{\varepsilon,m=0,\alpha}(\vec{r})$
having  $J_0(k_\alpha r)$ for its up-spin component and  $\psi_{\varepsilon,m=-1,\alpha}(\vec{r})$
having $J_0(k_\alpha r)$ for its down-spin component make nonzero contributions to
the matrix elements of the $t$-matrix,
$\bra{\psi_{\varepsilon m^\prime \alpha^\prime}} \mathbf{t}\ket{\psi_{\varepsilon m \alpha}}$,
since $J_m(0)$ vanishes for $m>0$.

We aim at deriving the general expression of the impurity resistivity when the $t$-matrix is
given by Eq.~\eqref{eq:t_delta}. First, we note that by using Eq.~\eqref{eq:lipp_r}, the
scattering coefficients $C(m\alpha,m^\prime\alpha^\prime)$ for $m$ and $m^\prime$ equal to 
0 or $-1$ are given by
\begin{eqnarray}
C(0\alpha,0\alpha^\prime)&=&\frac{1}{i} \delta_{\alpha\alpha^\prime}
-\frac{\sqrt{k_\alpha k_{\alpha^\prime}}}{2k_M} t_{\uparrow\uparrow} \nonumber \\
C(-1\alpha,-1\alpha^\prime)&=&i \delta_{\alpha\alpha^\prime}
+\frac{\sqrt{k_\alpha k_{\alpha^\prime}}}{2k_M} s(\alpha) s(\alpha^\prime) 
 t_{\downarrow\downarrow} \nonumber \\
C(0\alpha,-1\alpha^\prime)&=& \frac{\sqrt{k_\alpha k_{\alpha^\prime}}}{2i k_M} s(\alpha^\prime) 
 t_{\downarrow\uparrow}\nonumber\\
C(-1\alpha,0\alpha^\prime)&=& \frac{\sqrt{k_\alpha k_{\alpha^\prime}}}{2i k_M} s(\alpha) 
 t_{\uparrow\downarrow},\label{eq:c00} 
\end{eqnarray}
where $s(\alpha)$ is defined by $s(\pm 1)=\mp 1$ and the energy argument for the
$t$-matrix is omitted for simplicity.
For $m$ and $m^\prime$ larger than 0 or smaller than $-1$, we have
\begin{equation}
C(m\alpha,m^\prime\alpha^\prime)=\delta_{mm^\prime}\delta_{\alpha\alpha^\prime}
e^{-i\left(m+\frac{1}{2}\right)\pi}.\label{eq:cc2}
\end{equation}

The expression of the impurity resistivity can be obtained by inserting Eqs.~\eqref{eq:c00}
and \eqref{eq:cc2} into Eq.~\eqref{eq:der7}, where
twelve pairs of $(m,m^\prime)$ make a non-vanishing contribution to the resistivity.
After a lengthy calculation, one obtains
\begin{equation}
\rho_{\gamma \gamma}
=\frac{\hbar k^{2}_\mathrm{FM}}{2\pi S n_e^2 e^2}
\Big[\sum_{\sigma,\sigma^\prime} |t_{\sigma \sigma^\prime}|^2
-\left(\frac{k_\mathrm{so}}{k_\mathrm{FM}}\right)^2\left(M\mp N\right)\Big],
\label{eq:ssct1}
\end{equation}
with
\begin{eqnarray}
M&\equiv& \mathrm{Re}\Big(t^*_{\uparrow\uparrow} t_{\downarrow\downarrow}\Big)
+\frac{1}{2}(|t_{\uparrow\downarrow}|^2-|t_{\downarrow\uparrow}|^2)^2,\label{eq:ssct2}\\
N&\equiv&\mathrm{Re}\Big(t^*_{\uparrow\downarrow} t_{\downarrow\uparrow}\Big)\nonumber\\
&-&\frac{1}{2}\mathrm{Re}\Big[
(t^*_{\uparrow\uparrow}-t^*_{\downarrow\downarrow})\ t_{\downarrow\uparrow}
-( t_{\uparrow\uparrow}-    t_{\downarrow\downarrow})\ t^*_{\uparrow\downarrow}\Big]^2,
\label{eq:ssct3}
\end{eqnarray}
where the $t$-matrix should be evaluated at the Fermi energy and the negative and
positive signs in Eq.~\eqref{eq:ssct1} correspond to $\rho_{xx}$ and $\rho_{yy}$,
respectively.
In deriving the above equations, we have used the general relation
for the $t$-matrix (optical theorem), $\mathbf{t}-\mathbf{t}^\dagger=-i\ \mathbf{t}\mathbf{t}^\dagger$,
implying in the case of a 2$\times$2 matrix that
\begin{eqnarray}
\mathrm{Im}\ t_{\uparrow\uparrow}&=&-\frac{1}{2}
\left(|t_{\uparrow\uparrow}|^2+|t_{\downarrow\uparrow}|^2\right),\nonumber\\
\mathrm{Im}\ t_{\downarrow\downarrow}&=&-\frac{1}{2}
\left(|t_{\downarrow\downarrow}|^2+|t_{\uparrow\downarrow}|^2\right),\nonumber\\
t^*_{\uparrow\downarrow}-  t_{\downarrow\uparrow}
&=&i \left( t_{\uparrow\uparrow} t^*_{\uparrow\downarrow}
+t^*_{\downarrow\downarrow} t_{\downarrow\uparrow}\right).\nonumber
\end{eqnarray}
Because of the above relations, $M$ and $N$ in Eqs.~\eqref{eq:ssct2} and \eqref{eq:ssct3}
may be expressed in many apparently different but equivalent ways.
 
\section{Results and discussions}
\label{sec:Results}

\begin{figure*}
\hspace{-10mm}
\begin{minipage}[b]{0.4\textwidth}
\includegraphics[width=9cm]{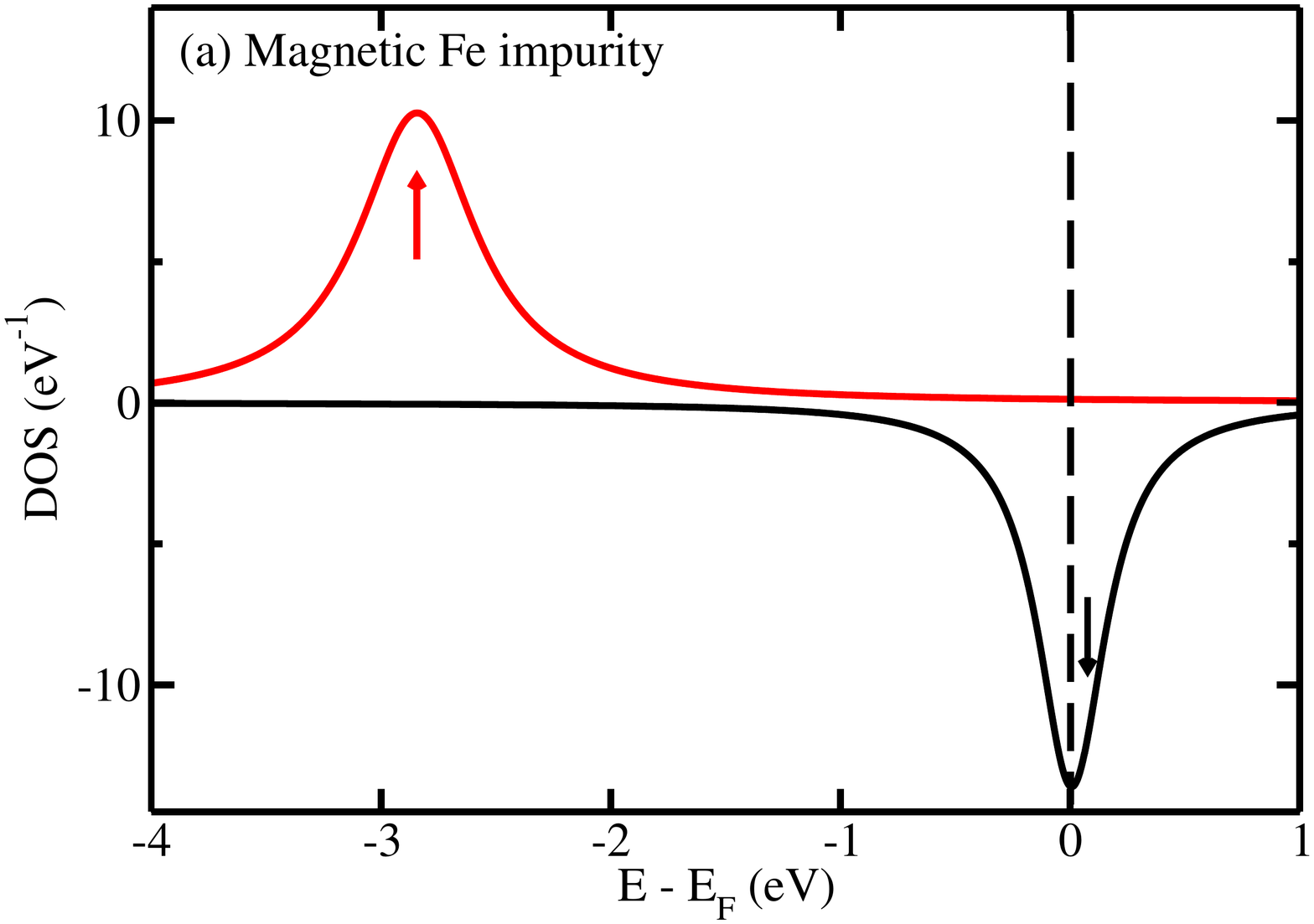}
\end{minipage}
\quad\quad\quad
\begin{minipage}[b]{0.4\textwidth}
\includegraphics[width=9cm]{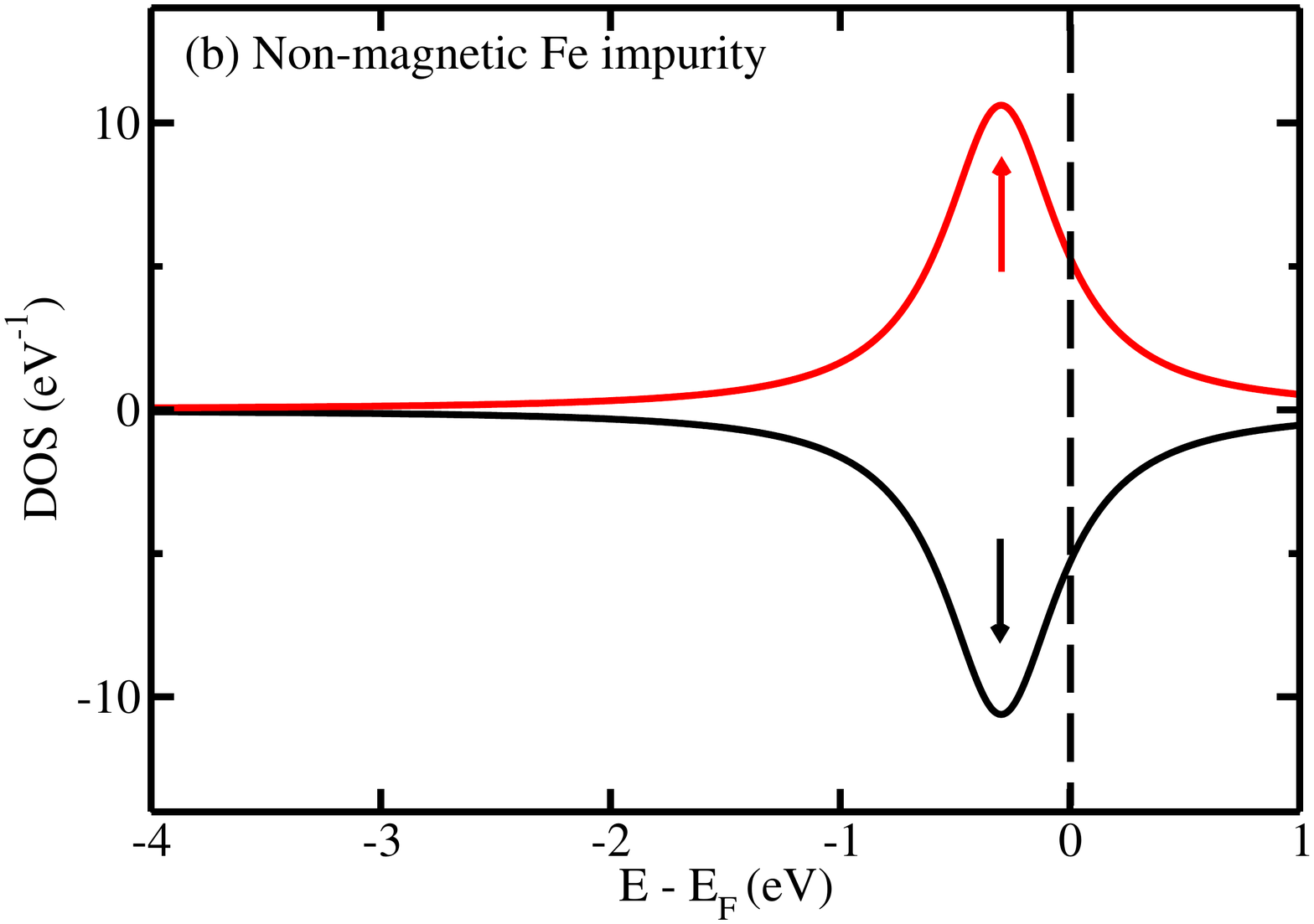}
\end{minipage}
\caption{Local Density of States of an Fe adatom deposited on a Au(111) surface described 
            by a Lorentzian model wherein the broadening is induced by hybridization effects among the 
            electronic states of the impurity with those of the substrate. Two cases are considered, a 
            magnetic (a) versus a non magnetic impurity (b). After defining the phase shifts at the Fermi 
            energy in the magnetic case, the phase shifts in the non-magnetic
            case are derived considering the same charge for both type of impurities.}
\label{density_of_states}
\end{figure*}

As aforementioned, we investigate the example of Fe-adatoms on Au(111) 
surface with an area $S = 1$ m$^{2}$ (unit area), considering the scattering 
of the Shockley surface states of Au(111) at the adatom. 
We assume that the impurity has no on-site SO coupling term. Furthermore, 
we will consider only cylindrical potentials. In the case where the magnetic moment of the impurity
is oriented in the $z$ direction, the $t$-matrix can be expressed by Eqs.~\eqref{eq:t_diagonal} and \eqref{eq:t_cyl}. 
Since the characterisic wave lengths of the Rashba electrons at $\varepsilon_\mathrm{F}$ 
corresponding to $k_{F+}$ and $k_{F-}$ are much larger than the impurity size, we can proceed with the so-called 
s-wave approximation (see section~\ref{subsec_swave}). Indeed the wavelengths $\lambda_{\mathrm{F}+}= 2\pi/0.192 = 32.7$~\AA, $\lambda_{\mathrm{F}-}= 
2\pi/0.167 =37.6$~\AA ~are large considering $k_{\mathrm{so}-}= -0.0125$~\AA$^{-1}$ ~with parameters taken from 
Ref.~\cite{Reinert:01}.  The advantages of this approximation are the fast numerical 
evaluation of the scattering coefficients and an easy tracking of the impact of scattering on the resistivity. 
The connection to the phase shift will be needed in the 
up-coming discussion and is given via: $t_{\sigma\sigma}(\varepsilon) =
 i\hspace{1mm} (e^{2i\delta_0(\varepsilon,\sigma)}-1)$, which are the diagonal elements of the $t$-matrix in spin 
space (see Eq.~\ref{eq:t_diagonal}).

The s-wave approximation has been used numerous times for the interpretation of 
scanning tunneling microscopy based 
experiments\cite{Fiete_RMP2003,Crampin_PRL1994,Crampin_PRB1996,Harbury_PRB1996}. This was done in the context of standing 
waves on Cu(111) surface \cite{Crommie:1993} or Au(111) surface 
\cite{Hasegawa_PRL1993} and confined electronic states  in corrals of Fe 
or Co adatoms deposited on Cu(111) surface\cite{Crommie_Science1993,Heller_Nature1994}. 
For Fe adatoms on Cu(111) surface, good fits to the experimental features were obtained with 
a phase shift of $\pi/2$ but a better agreement was found   with a phase shift of $i\infty$, which would correspond to 
maximaly absorbing adatoms (black dots)\cite{Heller_Nature1994}. In the latter case, 
the overall scattering amplitude reduces by a factor of 2 compared to a phase shift of $\pi/2$.

We follow a description similar to the one of Heller et al.\cite{Heller_Nature1994} but considering 
a phase shift of $\pi/2$ in the minority-spin channel. The majority-spin channel is considered fully occupied in the magnetic case and thus 
the corresponding phase shift is set to $\pi$. These assumptions were used in Ref.\cite{Samir:2012} and 
are conforted by our ab-initio simulations 
 based on density functional theory as implemented in the Korringa-Kohn-Rostoker Green function method\cite{KKR}. 
 From these calculations we learned that the easy axis of the Fe magnetic moment is out-of-plane.
The adatom local density of states (LDOS) is characterized by a resonance close to $\varepsilon_\mathrm{F}$
in the minority-spin channel and the exchange splitting between the majority-spin and minority-spin resonances 
is about 2.8 eV. The broadening of the resonances in the magnetic case is 0.6 eV in the majority-spin channel and 0.4 eV 
in the minority spin channel while in the non-magnetic case, it is considered to be 0.6 eV. Therefore, we assume that the LDOS considered in our model 
follows  the Lorentzian shapes depicted in  FIG.~\ref{density_of_states}(a). We note that in our
scheme based on linear response theory, only the phase shift at the Fermi energy is essential.

In order to evaluate the impact of magnetism on the residual resistivity, we consider a non-magnetic Fe adatom and 
use charge conservation in order to guess the appropriate parameters. The 
spin-dependent charge $N_{\sigma}$ of the impurity is given by 
$\frac{1}{\pi}\ \delta_0(\varepsilon_{F},\sigma)$ with the Friedel sum rule. Charge conservation imposes then that 
in the non-magnetic case $\delta_0(\varepsilon_{F},\uparrow) = \delta_0(\varepsilon_{F},\downarrow) = \frac{3\pi}{4}$, 
which leads to the LDOS plotted in FIG.~\ref{density_of_states}(b).

Considering the approximations mentioned above one can 
investigate the residual resistivity for both cases: magnetic and non-magnetic Fe adatoms. 
To start our analysis, we consider a magnetic moment pointing perpendicular 
to the surface. Also to make our study general, we explore different SO coupling strengths, which 
then would correspond to the deposition of the impurities on different substrates. 
To make this type of investigations consistent with each other, 
the energy of the highest occupied state of the Rashba electron gas measured from the bottom 
of the energy dispersion curve, $\varepsilon_F^{\mathrm{Rashba}} =  \varepsilon_F + \frac{\hbar^2}{2m^*}k_{\mathrm{so}}^2$, is set to a constant, 480 meV for 
the case of Au(111) surface state characterized by an effective mass $m^* = 0.255 m_e$~\cite{Reinert:01}. By changing $k_{\mathrm{so}}$, $\varepsilon_F$ is modified 
such that $\varepsilon_F^{\mathrm{Rashba}}$ does not vary.

By inserting
$t_{\uparrow\uparrow}=i(e^{i \delta_0(\varepsilon_\mathrm{F},\uparrow)}-1)$,
$t_{\downarrow\downarrow}=i(e^{i \delta_0(\varepsilon_\mathrm{F},\downarrow)}-1)$, and
$t_{\uparrow\downarrow}=t_{\downarrow\uparrow}=0$ into Eq.~\eqref{eq:ssct1}, we obtain as the 
resistivity induced by a magnetic adatom with its magnetic moment pointing to the normal direction,
\begin{eqnarray}
\rho_{\gamma \gamma}&=&\frac{2\hbar k^2_\mathrm{FM}}{\pi S n^2_e e^2}
\Bigg[\sin^2(\delta_\uparrow) + \sin^2(\delta_\downarrow)
\nonumber\\
&-&\left(\frac{k_\mathrm{so}}{k_\mathrm{FM}}\right)^2
\sin(\delta_\uparrow)\sin(\delta_\downarrow)\cos(\delta_\uparrow-\delta_\downarrow))\Bigg],
\label{eq:ssct4}
\end{eqnarray} 
where $\delta_\uparrow$ and $\delta_\downarrow$ are abbreviations of  $\delta_0(\varepsilon_\mathrm{F},\uparrow)$
and $\delta_0(\varepsilon_\mathrm{F},\downarrow)$, respectively. Thus, for the present non-magnetic
adatom with $\delta_\uparrow=\delta_\downarrow=\frac{3\pi}{4}$,
\begin{equation}
\rho_{\gamma \gamma}=\frac{2\hbar k^2_\mathrm{FM}}{\pi S n^2_e e^2}
\left[1-\frac{1}{2} \left(\frac{k_\mathrm{so}}{k_\mathrm{FM}}\right)^2\right],\label{eq:ssct5}
\end{equation}
while for the present magnetic adatom with $\delta_\uparrow=\pi$ and
$\delta_\downarrow=\pi/2$, we simply have
\begin{equation}
\rho_{\gamma \gamma}=\frac{2\hbar k^2_\mathrm{FM}}{\pi S n^2_e e^2},\label{eq:ssct6}
\end{equation}
where $k_\mathrm{FM}$ and $n_e$ are related to $\varepsilon_{F}^{\mathrm{Rashba}}$, which is
kept constant in the numerical calculation, by
$k^2_\mathrm{FM}=2m^* \varepsilon_{F}^{\mathrm{Rashba}}/\hbar^2$
and $n_{e}= \frac{m^*\varepsilon_{F}^{\mathrm{Rashba}}}{\pi\hbar^2} +
\frac{k_{\mathrm{so}}^2}{2\pi}$. The latter leads to a quadratic decrease of the prefactor $\frac{2\hbar k^2_\mathrm{FM}}{\pi S n^2_e e^2}$ and thus of the 
residual resistivity with 
respect to $k_{\mathrm{so}}^2$. This simply indicates that the more available electrons, 
the more conducting the system becomes.

The intriguing dependence of $n_e$ on the SO coupling strength can be traced 
back to the particular behavior of the density of states of the Rashba 
electron gas, which is characterized by two regimes induced by SO and defined by the 
two regions of the energy dispersion curve that show a crossing at $k = 0$ 
(see Eq.~\eqref{energy_dispersion}). At energies below the crossing, the corresponding density 
of states follows a quasi one-dimensional behavior where a van Hove singularity 
occurs at the bottom of the bands. Above the crossing,  the density of states is a  constant as expected 
for a 2D electron gas. By increasing the SO coupling strength, the quasi one-dimensional 
region becomes larger in order to keep $\varepsilon_F^{\mathrm{Rashba}}$ constant, 
which leads to the quadratic dependence of $n_e$ on $k_{\mathrm{so}}$ and explains the 
strong drop of the residual resistivity when increasing the SO coupling strength.

\begin{figure}
\centering
\includegraphics[width=8cm , trim= 1 1 1 1, clip=true,angle =0]{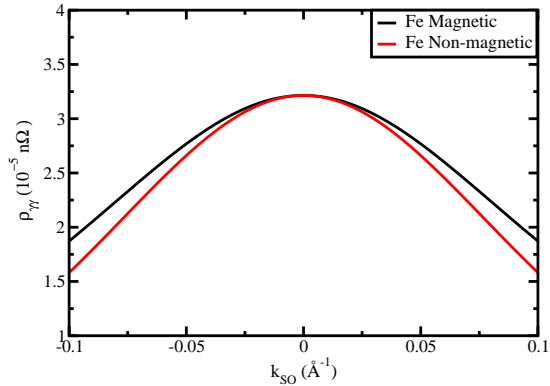}
 \caption{ Evolution of the diagonal components of the resistivity tensor as a 
                function of the spin-orbit wave vector length ($k_\mathrm{so}$) for a magnetic 
                and a non-magnetic Fe impurity.} 
  \label{Evolution}
\end{figure}

The latter can be observed in Fig.~\ref{Evolution}, where the longitudinal residual resistivity as function of SO is depicted.
The transveral  residual resistivity is not shown since it is zero for the two cases considered: 
magnetic (out-of-plane moment) and non-magnetic adatoms.
 Interestingly, magnetism and SO coupling strength have opposite impact on the residual resistivity. This holds 
for spin-dependent phase shifts that 
conserve the number of electrons $N$ of the impurity after spin-polarization. Indeed,
 as may be seen from Eq.~\eqref{eq:ssct4}, 
in contrast to magnetism, SO coupling tends to decrease the resistivity. The largest resistivity 
is found when the SO interaction is switched off, which would correspond to the case of Cu(111) surface. Here 
the residual resistivity is independent from the magnetic nature of the adatom as can be deduced from Eq.\eqref{eq:ssct4}
\begin{equation}
\rho_{\gamma \gamma} = \frac{ 4\hbar}{S n_\mathrm{e} e^{2}} 
\Big{[}\sin^2(\delta_\uparrow) + \sin^2(\delta_\downarrow)\Big{]},
\label{eq:res_nocsoc_r2r}
\end{equation}


\begin{figure}%
\hspace{-0.75cm}
\includegraphics[width= 9.3cm]{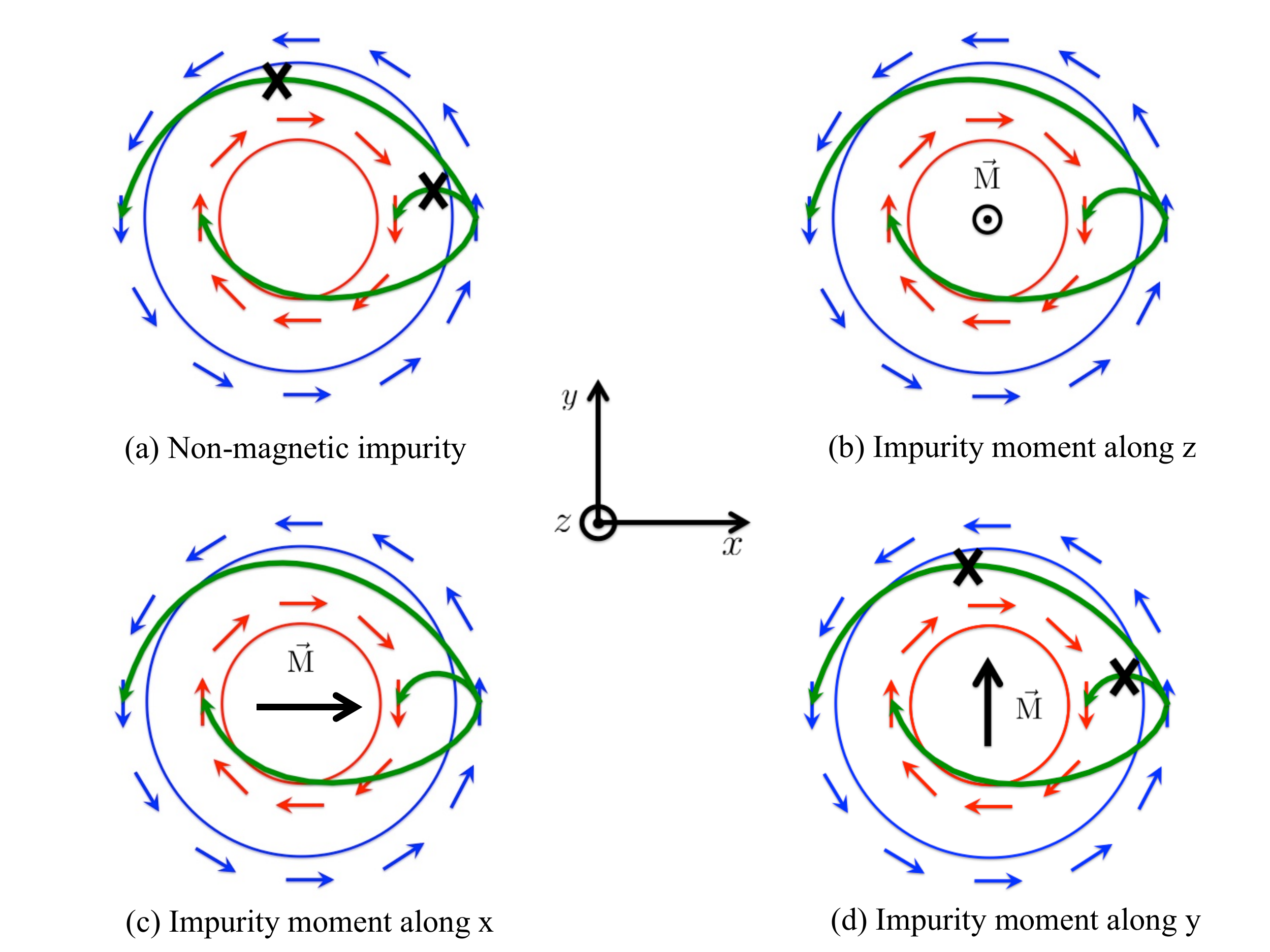}
\caption{Fermi surfaces scattering processes of Rashba electrons at a non-magnetic impurity 
              (a), magnetic impurity with an out-of-plane magnetic moment (b),
              in-plane magnetic moment pointing along the $x$-direction (c) and
              along the $y$-direction (d). The transitions between circles with different 
              colors are interband transitions, while transitions between circles with the 
              same color are intraband transitions. The green arrows indicate the connection between 
              the initial and final state. The cosses indicate prohibited scattering processes, while the 
              black arrows at the center of the Fermi contours represent the direction of the impurity magnetic moment.}%
\label{Fermi_surface}%
\end{figure}

Now, we address the dependence of the residual resistivity on the magnetism of the impurity by analyzing the 
different scattering processes allowed at the Fermi surface. The possible elastic scattering processes can be found 
by evaluating the probability for an electron scattering from a state $\ket{\psi_{\vec{k}\alpha}}$ to a state
$\ket{ \psi_{ \vec{k}^{\prime} \alpha^{\prime} }}$:
\begin{equation}
P^{\alpha\alpha^\prime}_{\vec{k}\vec{k}^\prime} = \frac{2\pi}{\hbar} |\bra{\psi_{\vec{k^\prime}\alpha^\prime}} \mathbf{t} \ket
{\psi_{\vec{k}\alpha}}|^2\ \delta(\varepsilon_{\vec{k}\alpha}-\varepsilon_{\vec{k}^\prime\alpha^{\prime}}),
\end{equation}
where $\psi_{\vec{k}\alpha}$
are given by Eq.~\eqref{Rashba_eigen} and $\alpha$ is the band index. If the impurity is non-magnetic, 
the diagonal elements of the $t$-matrix in spin space are equal: $t_{\uparrow\uparrow} = t_{\downarrow\downarrow} = t$. In this case, 
the electron scattering probabilities are given by
\begin{equation}
P^{\alpha\alpha^\prime}_{\vec{k}\vec{k}^\prime} =
\frac{\pi}{2\hbar}|t|^2(1+{\alpha}{\alpha^\prime}\cos{(\phi_{\vec{k}}-\phi_{\vec{k}^\prime})})
\ \delta(\varepsilon_{\vec{k}\alpha}-\varepsilon_{\vec{k}^\prime\alpha^{\prime}}), 
\end{equation}
where $\alpha\alpha^\prime$ equals 1 for intra-bad scattering transitions ($\alpha = \alpha^\prime$) or -1 for inter-band 
transitions ($\alpha \neq \alpha^\prime$).
This equation shows that inter-band and intra-band transitions 
flipping the spin are not allowed since in these cases, 
$\phi_{\vec{k}}-\phi_{\vec{k}^\prime} = \pi$ with $\alpha\alpha^\prime = 1$ and  
$\phi_{\vec{k}}-\phi_{\vec{k}^\prime} = 0$ with $\alpha\alpha^\prime = -1$, respectively (see Fig.~\ref{Fermi_surface} (a)). 

In the case of a magnetic impurity with a moment 
perpendicular to the surface, the $t$-matrix is given by Eq.~\ref{eq:t_diagonal} and all transitions are allowed, even those leading to a spin-flip, 
as depicted in Fig.~\ref{Fermi_surface} (b). Here the electron scattering probabilities 
\begin{equation}
P^{\alpha\alpha^\prime}_{\vec{k}\vec{k}^\prime} = \frac{2\pi}{\hbar}|t_{\uparrow \uparrow} +\alpha\alpha^\prime 
t_{\downarrow \downarrow}e^{i(\phi_{\vec{k}}-\phi_{\vec{k}^\prime})}|^2\ \delta(\varepsilon_{\vec{k}\alpha}-\varepsilon_{\vec{k}^\prime\alpha^{\prime}})
\end{equation}
which is different from zero independently from the value of $\phi_{\vec{k}}-\phi_{\vec{k}^\prime}$. This is due 
to the magnetic moment of the impurity which breaks the time-reversal symmetry. 
Thus, there are more scattering possibilities than in the non-magnetic case, and therefore for magnetic impurities a higher residual resistivity is expected
as to non-magnetic impurities in-line with Fig.~\ref{Evolution}. 

\begin{figure*}
\centering
\includegraphics[trim=0.8cm 10.5cm 0.1cm 0cm, clip=true,width=15cm,scale=1.0, angle=0]{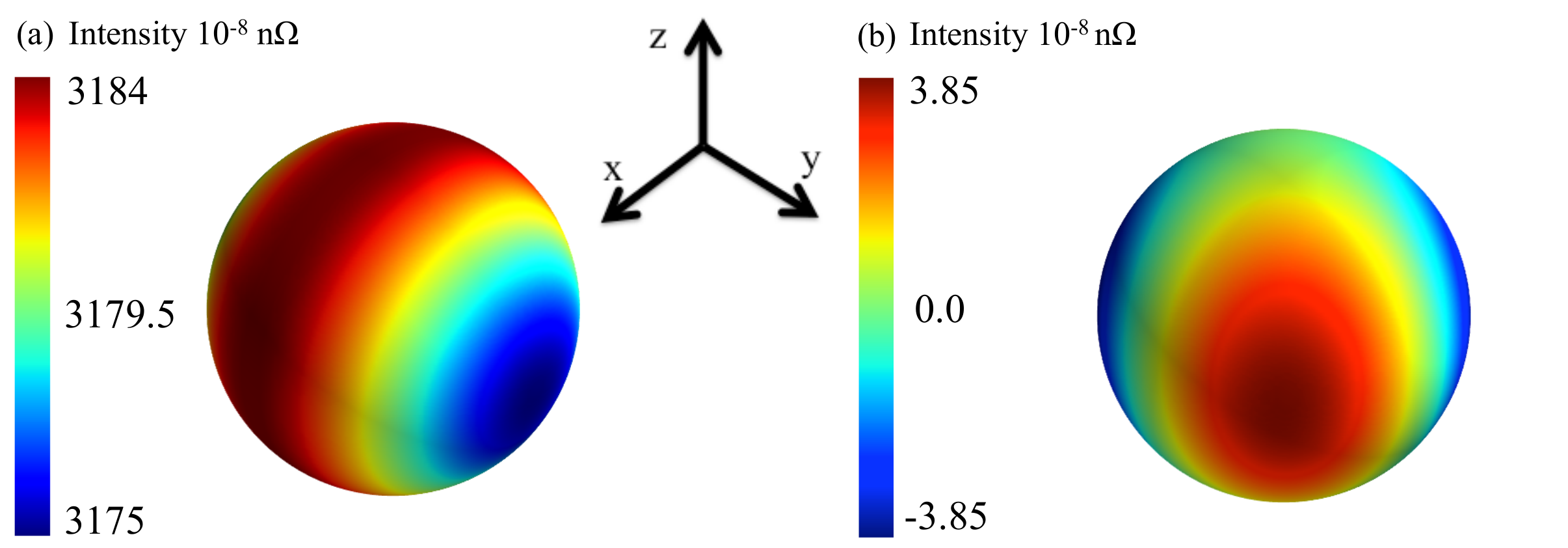}
\quad
\begin{minipage}[b]{0.4\textwidth}
\hspace{-10mm}
\includegraphics[width=8cm]{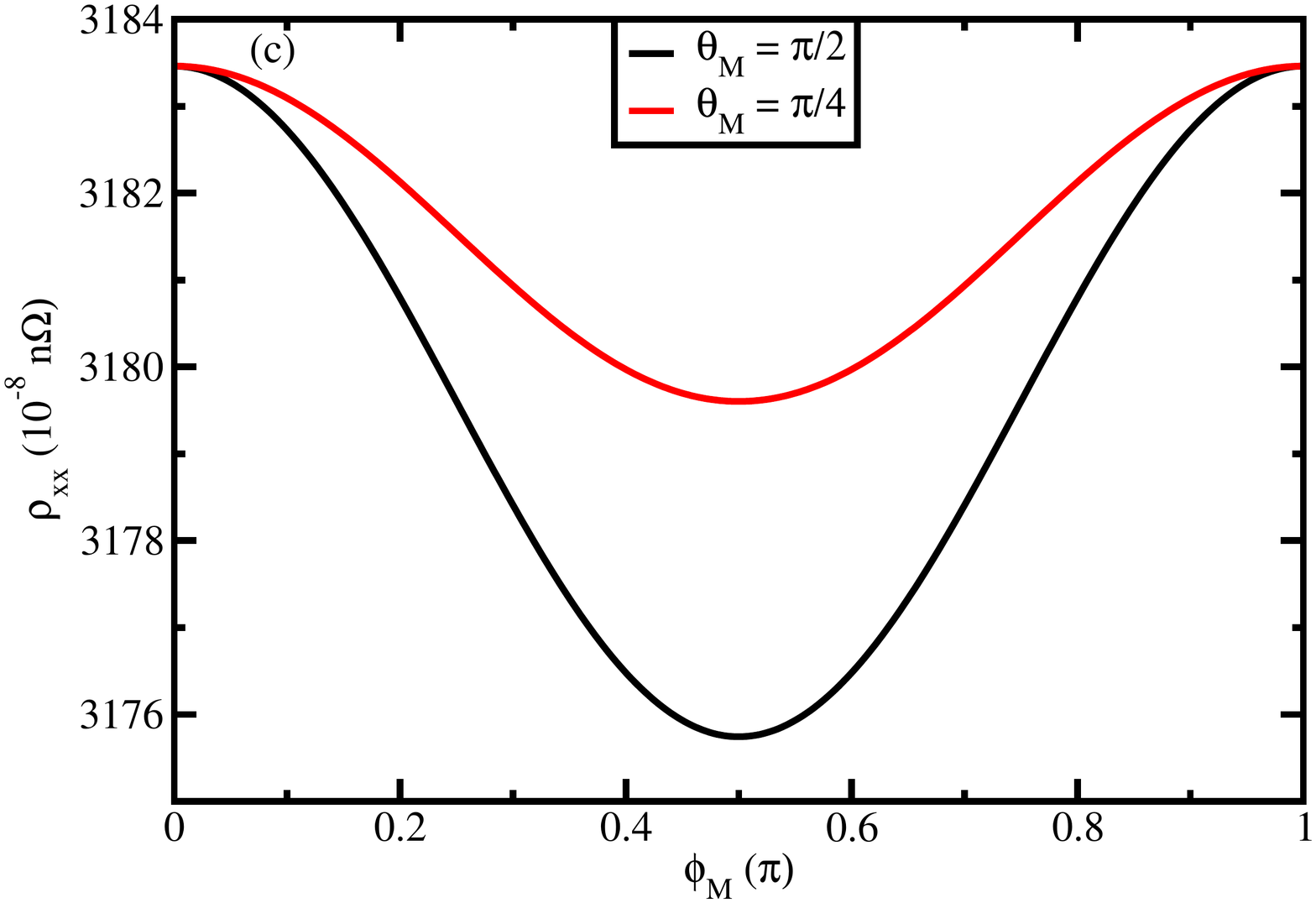}
\end{minipage}
\quad\quad\quad 
\begin{minipage}[b]{0.4\textwidth}
\hspace{-10mm}
\includegraphics[width=8cm]{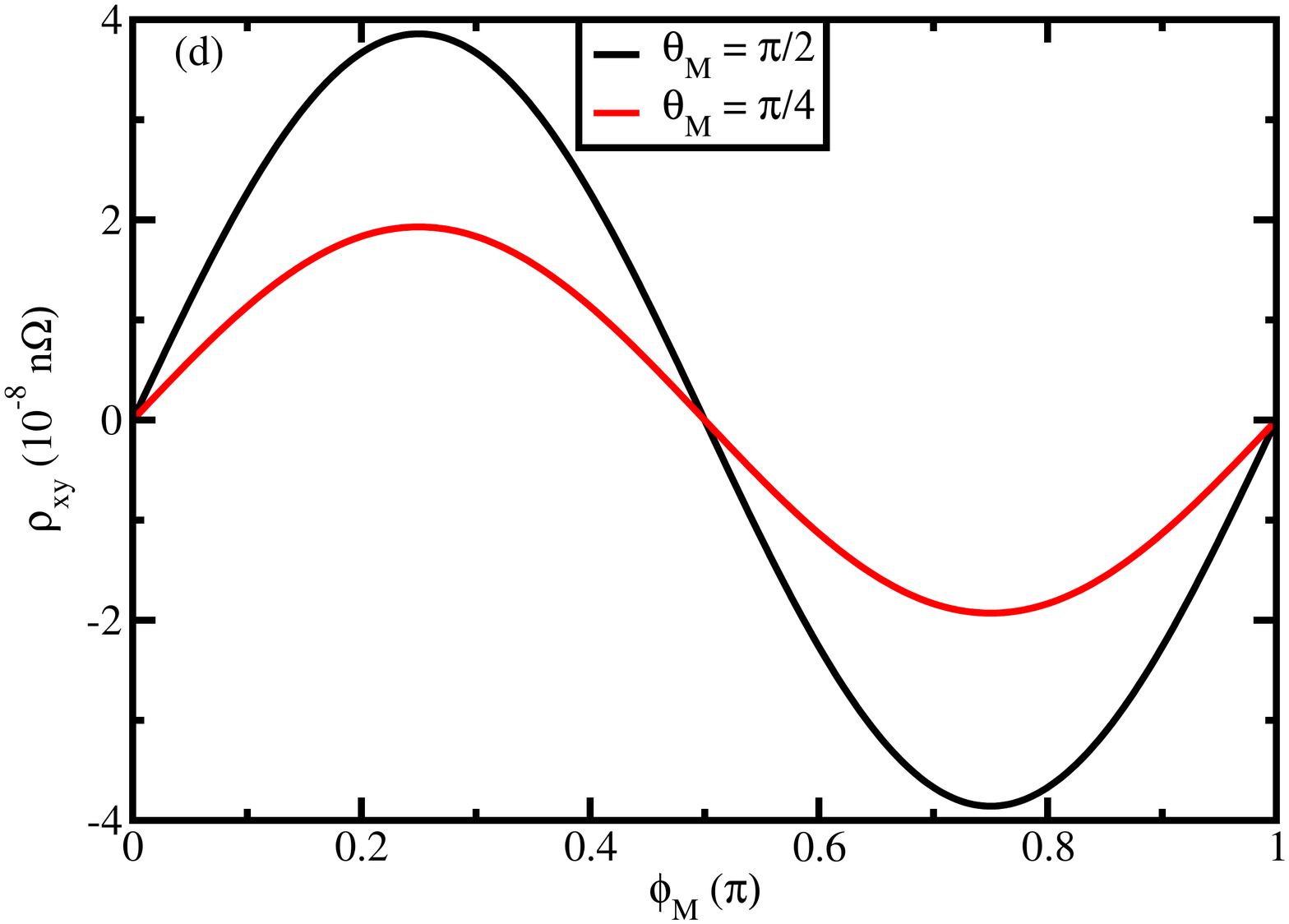}
\end{minipage}
\caption{Evolution of the components of the residual resistivity tensor as function of the orientation of the magnetic
 moment in all 4$\pi$ spatial directions: (a) longitudinal component ($\rho_{xx}$), 
(b) transversal component ($\rho_{xy}$). Every point on the sphere corresponds to a given 
orientation of the magnetic moment. 
 Evolution of the residual resistivity tensor components while 
changing only $\phi_{\vec{\mathrm{M}}}$ when the magnetic moment is pointing in-plane along the $x$-direction: (c) longitudinal 
and (d) transversal component. Here we plotted the following cases $\theta_{\vec{\mathrm{M}}} = 
\frac{\pi}{2}$ (black curve) and $\theta_{\vec{\mathrm{M}}} = \frac{\pi}{4}$ (red curve).}
    \label{3D_res}%
\end{figure*}

Up to now, the magnetic moment was considered perpendicular to the surface plane. 
To generalize our study, we explore the impact of arbitrary orientations, 
$\hat{e}_{\vec{\mathrm{M}}}$, of the impurity moment, $\vec{\mathrm{M}}$, on 
the residual resistivity. The resistivity is a tensor, and contrary to the case of an 
out-of-plane magnetic orientation, its off-diagonal elements become finite for
 arbitrary magnetization directions giving rise to the AMR and PHE. 
To tackle this problem, we rotate the impurity magnetic moment pointing initially normal to the surface 
plane, $\hat{e}_z$, by means of the conventional 
$3\times 3$ rotation matrices ${\cal R}\in \mathrm{SO(3)}$ by a polar angle $\theta_{\vec{\mathrm{M}}}$  between the direction of the 
magnetic moment and the $z$-axis and an azimuthal angle 
$\phi_{\vec{\mathrm{M}}}$, $\hat{e}_{\vec{\mathrm{M}}}={\cal R}_z(\phi_{\vec{\mathrm{M}}}){\cal R}_y(\theta_{\vec{\mathrm{M}}})\,\hat{e}_z$. 
This translates to a unitary transformation of the $t$-matrix in spin space   
$\mathbf{t}^\prime(\vec{r}, \vec{r}^{\,\prime}) = \mathbf{U}(\theta_{\vec{\mathrm{M}}}, \phi_{\vec{\mathrm{M}}})
\hspace{1mm} \mathbf{t}(\vec{r}, \vec{r}^{\,\prime})
\hspace{1mm}\mathbf{U}^{\dag}(\theta_{\vec{\mathrm{M}}},\phi_{\vec{\mathrm{M}}})$. 
$\mathbf{U}(\theta_{\vec{\mathrm{M}}}, \phi_{\vec{\mathrm{M}}})= \mathbf{U}({\cal R}_z
,\phi_{\vec{\mathrm{M}}})\mathbf{U}({\cal R}_y,\theta_{\vec{\mathrm{M}}})$ 
are the conventional rotation operators in SU(2) whose representation in terms of a $2\times2$ matrix is given by 
$\mathbf{U}({\cal R}_\gamma,\beta)=\cos(\beta/2)\otimes\unitmatr_2 -i\sin(\beta/2)\boldsymbol{\sigma}_\gamma$.   
The $t$-matrix for an arbitrary rotation 
angle of the moment can then 
be expressed in terms of the spin diagonal elements, $t_{\uparrow\uparrow}$ and $t_{\downarrow\downarrow}$, describing the out-of-plane moment 
(see Eq.~\ref{eq:t_diagonal}) as:
\begin{equation}
\mathbf{t} = \frac{1}{2}(t_{\uparrow \uparrow} 
+ t_{\downarrow \downarrow}) \otimes \unitmatr_2 + 
\frac{1}{2}(t_{\uparrow \uparrow} - t_{\downarrow \downarrow}) \ \vec{\sigma}\cdot\hat{e}_{\vec{\mathrm{M}}}.
\label{eq:t_rotated}
\end{equation}
For given  
values of  $\{\theta_{\vec{\mathrm{M}}}, \phi_{\vec{\mathrm{M}}}\}$  the matrix $\mathbf{t}^\prime(\vec{r}, \vec{r}^{\,\prime})$ 
might have non-zero off-diagonal components. We note that 
we chose to define the azimuthal angle with respect to the $x$-axis being the direction of 
the perturbing current.

The longitudinal and transversal components of the residual resistivity in the whole 
phase space of rotation angles is depicted in FIG.~\ref{3D_res} (a) and (b).  
The images exhibit a clearly visible angular dependence. The anisotropy of the resistivity is in the order of $10^{-8}$ n$\Omega$.
 In case of the longitudinal resistivity this anisotropy modifies the isotropic contribution of the longitudinal resistivity, which 
is in the order of $3.180\times10^{-5}$ n$\Omega$ by about $\pm 0.14~\%$. Since for 
the transversal resistivity the isotropic contribution is exactly zero, 
the anisotropy are given as absolute values. 

Now we turn to the analysis of the angular dependence 
of the resistivity anisotropy. To the best of our knowledge,
no phenomenological functional form for such a general 
angular dependence is available in the literature contrary to the case, for example, 
where the magnetization is lying in-plane~\cite{D.A.Thompson:1975}. 
Therefore to simplify our analysis, we 
focus first on the particular orientation of the magnetic moment 
along the $x$-direction. There, the $t$-matrix given by Eq.\ref{eq:t_rotated} is expressed as
\begin{equation}
\mathbf {t}^\prime=
\frac{1}{2} \left(
\begin{array}{rcl}
\-t_{\uparrow \uparrow}+\-t_{\downarrow \downarrow}
&\-t_{\uparrow \uparrow}-\-t_{\downarrow \downarrow} \\ 
\-t_{\uparrow \uparrow}-\-t_{\downarrow \downarrow}
&\-t_{\uparrow \uparrow}+\-t_{\downarrow \downarrow}\\
\end{array}
\right),\label{eq:t_x}
\end{equation}
where $t_{\uparrow \uparrow}$ and $t_{\downarrow \downarrow}$ are the upper and lower diagonal components
of the $t$-matrix when the magnetic moment points along the $z$-direction.
As we will discuss below, this gives rise to a non-zero off-diagonal
contribution in the resistivity tensor and contributes to the PHE even without spin-orbit contribution  at the impurity site. 
When the magnetic moment is in the surface plane ($\theta_{\vec{\mathrm{M}}} = \frac{\pi}{2}$),  FIG.~\ref{3D_res}(c) and (d) 
show the behavior of the diagonal and off-diagonal components of the resistivity tensor, respectively,
 as function of the azimuthal angle  $\phi_{\vec{\mathrm{M}}}$.
The trace of the resistivity tensor is conserved under these azimuthal rotations ($\rho_
{xx} + \rho_{yy} =\ \mathrm{constant}, \forall \hspace{1mm} \phi_{\vec{\mathrm{M}}}$) while the off-diagonal components 
are related by $\rho_{xy} = \rho_{yx} $. 

The diagonal components of the resistivity tensor (FIG.~\ref{3D_res}(c)) can be fitted with 
the AMR functional form given by Thompson et al.\cite{D.A.Thompson:1975}: $\rho_{\gamma\gamma}
 = \rho_{\perp} + (\rho_{\parallel}-\rho_{\perp})\ \cos^2(\phi_{\vec{\mathrm{M}}})$, where $\rho_{\parallel}$ and $\rho_{\perp}$
define the residual resistivities when the moment is respectively parallel
 and perpendicular to the current. In our particular case, the current
 is pointing along the $x$-direction, thus $\rho_{\parallel} = \rho_{xx}
(\phi_{\vec{\mathrm{M}}}=0)$ and $\rho_{\perp} = \rho_{xx}(\phi_{\vec{\mathrm{M}}}=\frac{\pi}{2})$.
It turns out that $(\rho_{\parallel}-\rho_{\perp})$, 
 {\it i.e.} the maximal value of the AMR, is a positive
quantity as expected for a Rashba electron gas \cite{Trushin2009}. This can be explained by analyzing the different 
scattering processes on the Fermi surface when the impurity magnetic moment is in-plane (see FIGs.~\ref{Fermi_surface}(c) and (d)). 
As done previously, the idea 
is to evaluate the scattering probabilities for an arbitrary rotation of the magnetic moment (see Eq.\ref{eq:t_rotated}). Here we provide 
the results obtained for scattering processes from $\phi_{\vec{k}} = 0$ to $\phi_{\vec{k}^\prime} = 0$ or $\pi$. 
If $\phi_{\vec{k}^\prime} - \phi_{\vec{k}} = 0$ only inter-band transitions, i.e. $\alpha \neq \alpha^\prime$, can contribute: 
\begin{equation}
P^{\alpha\alpha^\prime}_{\vec{k}\vec{k}^\prime} = \frac{2\pi}{\hbar}|t_{\uparrow \uparrow} - t_{\downarrow \downarrow}|^2
(\cos^2{\theta_{\vec{\mathrm{M}}}}+ \sin^2{\theta_{\vec{\mathrm{M}}}}\cos^2{\phi_{\vec{\mathrm{M}}}})\ \delta(\varepsilon_{\vec{k}\alpha}-\varepsilon_{\vec{k}^\prime\alpha^{\prime}}),
\end{equation}
which is zero if the moment points along the $y$-direction. This is the same result obtained for intra-band scattering probability, 
$P^{\alpha\alpha}_{\vec{k}\vec{k}^\prime}$,  
when $\phi_{\vec{k}^\prime} - \phi_{\vec{k}} = \pi$.  For the later angle configuration, the inter-band scattering probability is non-zero 
independently from the rotation angle of the moment:
\begin{equation}
P^{\alpha\alpha^\prime}_{\vec{k}\vec{k}^\prime} = \frac{2\pi}{\hbar}
|t_{\uparrow \uparrow} + t_{\downarrow \downarrow}+\alpha(t_{\uparrow \uparrow} - t_{\downarrow \downarrow})\sin{\theta_{\vec{\mathrm{M}}}}\sin{\phi_{\vec{\mathrm{M}}}}|^2 
\ \delta(\varepsilon_{\vec{k}\alpha}-\varepsilon_{\vec{k}^\prime\alpha^{\prime}}).
\end{equation}

To summarize, when the magnetic moment points along the current direction ($x$-direction) the back-scattering is due to
inter-band and intra-band scattering. However when the magnetic moment is perpendicular to the current direction then  
the back-scattering is only originating from inter-band transitions, which induces a smaller residual resistivity and therefore
gives a positive maximal value of the AMR, i.e. $\rho_{\parallel} > \rho_{\perp}$. Similar scattering processes are possible when 
the moment points along the $x$-direction or the $z$-direction, which explains that the resistivities are the same for 
both magnetic orientations.

The off-diagonal components of the resistivity tensor (FIG.~\ref{3D_res}(d)) could be fitted with the PHE 
functional form \cite{D.A.Thompson:1975} $\rho_{\gamma\gamma^{\prime}} =  (\rho_{\parallel}-\rho_{\perp})
\ \cos(\phi_{\vec{\mathrm{M}}}) \sin(\phi_{\vec{\mathrm{M}}})$ We notice that for the considered polar 
angles $(\theta_{\vec{\mathrm{M}}} 
= \frac{\pi}{2}, \theta_{\vec{\mathrm{M}}} 
= \frac{\pi}{4})$, $\rho_{\gamma\gamma^{\prime}}$ changes sign when $\phi_{\vec{\mathrm{M}}}$ crosses $\frac{\pi}{2}$ (FIG.~\ref{3D_res}(d)).
This is accompanied by a direction switch of the Hall like electric field originating from the PHE.
The change in the sign of $\rho_{\gamma\gamma^{\prime}}$ reduces the energy dissipation $P$ 
given by Eq.~\ref{dissipation} since $\rho_{\gamma\gamma}$ is always positive.

Let us go back to the general case, where the magnetic moment points
 in arbitrary orientations. As mentioned earlier, a phenomenological functional form 
for the residual resistivity has not been proposed up to now. In Appendix~\ref{sec:Appendix_B}, we derive  
phenomenological functional forms for the residual resistivity tensor and show that
the longitudinal and transversal parts follow a simple angular dependence:
\begin{equation}
 \rho_{xx} = \rho_{\parallel}-(\rho_{\parallel}-\rho_{\perp}) \sin^{2}
(\phi_{\vec{\mathrm{M}}})\sin^{2}(\theta_{\vec{\mathrm{M}}}),\label{eq:rhoxx}   
\end{equation}
\begin{equation}
\rho_{xy} = (\rho_{\parallel}-\rho_{\perp})\cos(\phi_{\vec{\mathrm{M}}})\sin
(\phi_{\vec{\mathrm{M}}})\sin^{2}(\theta_{\vec{\mathrm{M}}}).
 \end{equation}
These equations describe perfectly the angular dependence plotted for instance in 
FIG.~\ref{3D_res}(c) and (d) with the polar angle $\theta_{\vec{\mathrm{M}}} 
= \frac{\pi}{2}$ (black curve) and $\theta_{\vec{\mathrm{M}}} = \frac{\pi}{4}$ (red curve).

Alternatively,  one may also derive the angular dependence of the diagonal components of the resistivity
tensor directly from Eq.~\eqref{eq:ssct1}. By applying the aforementioned unitary transformations
in spin space, Eq.~\ref{eq:t_rotated}, to the $t$-matrix and substituting its matrix elements into Eq.~\eqref{eq:ssct1} 
one yields the diagonal components of
the resistivity. 
 In the present case, the second terms of $M$ and $N$, which are proportional to the
fourth power of $t_{\sigma \sigma^\prime}$, vanish identically and one obtains
\begin{eqnarray}
\rho_{xx}&=&\frac{2\hbar k^2_\mathrm{FM}}{\pi S n^2_e e^2}
\Bigg[\sin^2(\delta_\uparrow) + \sin^2(\delta_\downarrow)
\nonumber\\
&-&\left(\frac{k_\mathrm{so}}{k_\mathrm{FM}}\right)^2
\sin(\delta_\uparrow)\sin(\delta_\downarrow)\cos(\delta_\uparrow-\delta_\downarrow)\nonumber\\
&-&\left(\frac{k_\mathrm{so}}{k_\mathrm{FM}}\right)^2\frac{1}{2}
\sin^2(\delta_\uparrow-\delta_\downarrow)
\sin^2\theta_{\vec{\mathrm{M}}} \sin^2\phi_{\vec{\mathrm{M}}}\Bigg], 
\label{eq:ssct7}
\end{eqnarray} 
where similarly to Eq.~\ref{eq:ssct4}, 
$\delta_{\uparrow}$ and $\delta_{\downarrow}$ are respectively abbreviations of $\delta_0({\epsilon_F,\uparrow})$ and 
$\delta_0(\epsilon_F,\downarrow)$. 
Thus, the magnitude of the AMR, i.e., $\rho_{\parallel}-\rho_{\perp}$ in Eq.~\eqref{eq:rhoxx} is given by
\begin{equation}
\rho_{\parallel}-\rho_{\perp}=\frac{\hbar k^2_\mathrm{so}}{\pi S n^2_e e^2}
\sin^2(\delta_\uparrow-\delta_\downarrow) \geq 0,
\end{equation}
indicating that the maximal value of the AMR occurs when the difference between the phase shifts
of both spin components becomes equal to $\pi/2$.

\section{Summary}
\label{sec:Summary}
Using linear response theory, we have derived a formulation of the tensor of the residual electrical resistivity for the 
particular case of a Rashba electron gas scattering at an impurity that can be magnetic and whose 
magnetic moment can point in any arbitrary direction. While the obtained form 
is general, we applied it to the case of an Fe impurity deposited at the Au(111) surface. We performed different types of 
studies and investigated the non-trivial impact of the strength of spin-orbit interaction of the substrate, the role of the magnetism of the impurity 
and of the orientation of the magnetic moment on the diagonal and off-diagonal elements of the resistivity tensor. 
 For instance, we found that after scattering, the planar hall effect and an 
anisotropic magnetoresistance occur even without incorporating the spin-orbit interaction at the 
impurity site if the orientation of the magnetic moment is not perpendicular to the surface.
Also an increase of the spin-orbit 
coupling strength induces a dramatic drop of the resistivity, which is related to a peculiar 
behavior of the electronic states of the Rashba electrons. Magnetism 
can increase the residual resistivity because of the opening of additional scattering 
channels, which were prohibited in the non-magnetic case. 
We derive analytically and generalize the usual phenomenological functional forms of the angular
 dependence of the resistivity tensor elements to the cases where 
the magnetization points in arbitrary directions. Finally, by switching-off
 the spin-orbit interaction, we find a simple formulation of the residual resistivity 
very close to the one given by Friedel for a 3 dimensional electron gas~\cite{Friedel:1958}.  

Our numerical results were obtained in the s-wave approximation involving a Rashba Hamiltonian and 
as discussed in the context of lifetime reduction of surface states by adatom scattering~\cite{Heers_PRB2012}, 
 it would be interesting to investigate the impact of realistic band structures computed from density functional theory 
on the residual resistivities and asses thereby the effect of other scattering channels besides the ones involving 
only surface states.

\begin{acknowledgments}
J.B. thanks M. dos Santos Dias and S.B. thanks Yuriy Mokrousov for fruitful discussions.
J.B. and  S.L. gratefully acknowledge
 funding under HGF YIG Program VH-NG-717 (Functional Nanoscale Structure and Probe Simulation Laboratory--Funsilab) 
and the DFG project LO 1659/5-1. S.B.  acknowledges  
funding under the DFG-SPP 1666 ``Topological Insulators: Materials -- 
Fundamental Properties -- Devices''.
The work of H.I. was supported by MEXT KAKENHI  No.\ 25110006
and by JSPS KAKENHI No. 24540328.

\end{acknowledgments}

\vspace{5ex}
\appendix
\section{Evaluation of the momentum operator matrix elements}
\label{sec:Appendix_A}
In this appendix we will calculate the matrix elements of the momentum operator between two quantum
states $i = (\varepsilon, m, \alpha)$ and $j = (\varepsilon+\hbar \omega, m^\prime, \alpha^\prime)$ 
in the region where the scattered wave function can be expressed by the asymptotic form Eq.~\eqref{eq:der1}
 and extract only the diverging terms  $\sim 1/\omega$ in the limit $\omega$ $\rightarrow$
 $0$. In cylindrical coordinates the leading components of the momentum operator behaves at
large distances as
\begin{equation}
\begin{split}
& v_{x} \sim  \frac{\hbar}{im^*} \cos 
\phi \hspace{1mm} \frac{\partial }{\partial r} + \frac{\hbar}{m^*} k_{\mathrm{so}}\ \sigma_{y}\\ 
 & v_{y} \sim \frac{\hbar}{im^*} \sin 
\phi \hspace{1mm} \frac{\partial }{\partial r} - \frac{\hbar}{m^*} k_{\mathrm{so}}\ \sigma_{x}.
\end{split}
\label{eq:der2}
\end{equation}
With this representation of the momentum operators, we found that the diverging terms arise from the  combinations
\begin{equation}
\begin{split}
&\Big\langle\psi_{\varepsilon+\hbar \omega, m^\prime, \alpha^{\prime}}^\mathrm{in,out}
\Big\vert m^*v_{x} \Big\vert\psi_{\varepsilon , m, \alpha}^\mathrm{in,out}\Big\rangle \\
&\sim \frac{\hbar(\delta_{m^\prime,m+1} +\delta_{m^\prime,m-1} )\delta_{\alpha
\alpha^\prime}}{2i \Delta k} \frac{k_{\mathrm{M}}}{k_{\alpha}}, 
\label{eq:asym1}
\end{split}
\end{equation}
where $\Delta k \equiv k_{\alpha}(\varepsilon+\hbar \omega) - k_{\alpha}(\varepsilon)
\sim 2m^*\omega/k_{\mathrm{M}}(\varepsilon)$ is the same for both bands ($\alpha=\pm$).

By combining Eq.~\eqref{eq:der1} and Eq.~\eqref{eq:asym1} we obtain the momentum matrix element
in the limit of $\omega \to 0$,
\begin{equation}
\bra{\varphi_{j}} m^*v_{x} \ket{\varphi_{i}}  \sim \frac{\hbar}{2i \Delta k} 
\frac{k_{\mathrm{M}}}{\sqrt{k_{\alpha}k_{\alpha}^{\prime}}} 
S^x(m\alpha, m^{\prime}\alpha^\prime, \varepsilon), 
\label{eq:der3B}
\end{equation}
where $S^x(m\alpha, m^{\prime}\alpha^\prime, \varepsilon)$ is given by
\begin{eqnarray}
S^x(m\alpha, m^{\prime}\alpha^\prime, \varepsilon) =
(\delta_{m\prime,m+1}+\delta_{m\prime,m-1})\delta_{\alpha \alpha^\prime} \nonumber \\
  + \sum_
{l\alpha^{\prime\prime}} C(m \alpha, l \alpha^{\prime\prime}) C^{*}(m^\prime \alpha^\prime, l+1 \alpha^{\prime\prime})
\nonumber \\ 
 + \sum_{l\alpha^{\prime\prime}} C(m \alpha, 
l \alpha^{\prime\prime}) C^{*}(m^\prime \alpha^\prime, l-1 \alpha^{\prime\prime}).\label{eq:der9B}
\end{eqnarray}
Analogously, the matrix element for the $y$ component of the momentum operator is given by
\begin{equation}
\bra{\varphi_{j}} m^*v_{y} \ket{\varphi_{i}}  \sim \frac{-\hbar}{2 \Delta k} 
\frac{k_{\mathrm{M}}}{\sqrt{k_{\alpha}k_{\alpha}^{\prime}}} 
S^y(m\alpha, m^{\prime}\alpha^\prime, \varepsilon), 
\label{eq:der3C}
\end{equation}
with $S^y$ defined by
\begin{eqnarray}
S^y(m\alpha, m^{\prime}\alpha^\prime, \varepsilon) =
(\delta_{m\prime,m+1}-\delta_{m\prime,m-1})\delta_{\alpha \alpha^\prime}  \nonumber \\
  + \sum_
{l\alpha^{\prime\prime}} C(m \alpha, l \alpha^{\prime\prime}) C^{*}(m^\prime \alpha^\prime, l+1 \alpha^{\prime\prime})
\nonumber  \\ 
 - \sum_{l\alpha^{\prime\prime}} C(m \alpha, 
l \alpha^{\prime\prime}) C^{*}(m^\prime \alpha^\prime, l-1 \alpha^{\prime\prime}). \label{eq:der9C}
\end{eqnarray}

\section{Phenomenological derivation of the functional forms}
\label{sec:Appendix_B}
\begin{figure}%
\hspace{-0.75cm}
\includegraphics[width= 9.3cm]{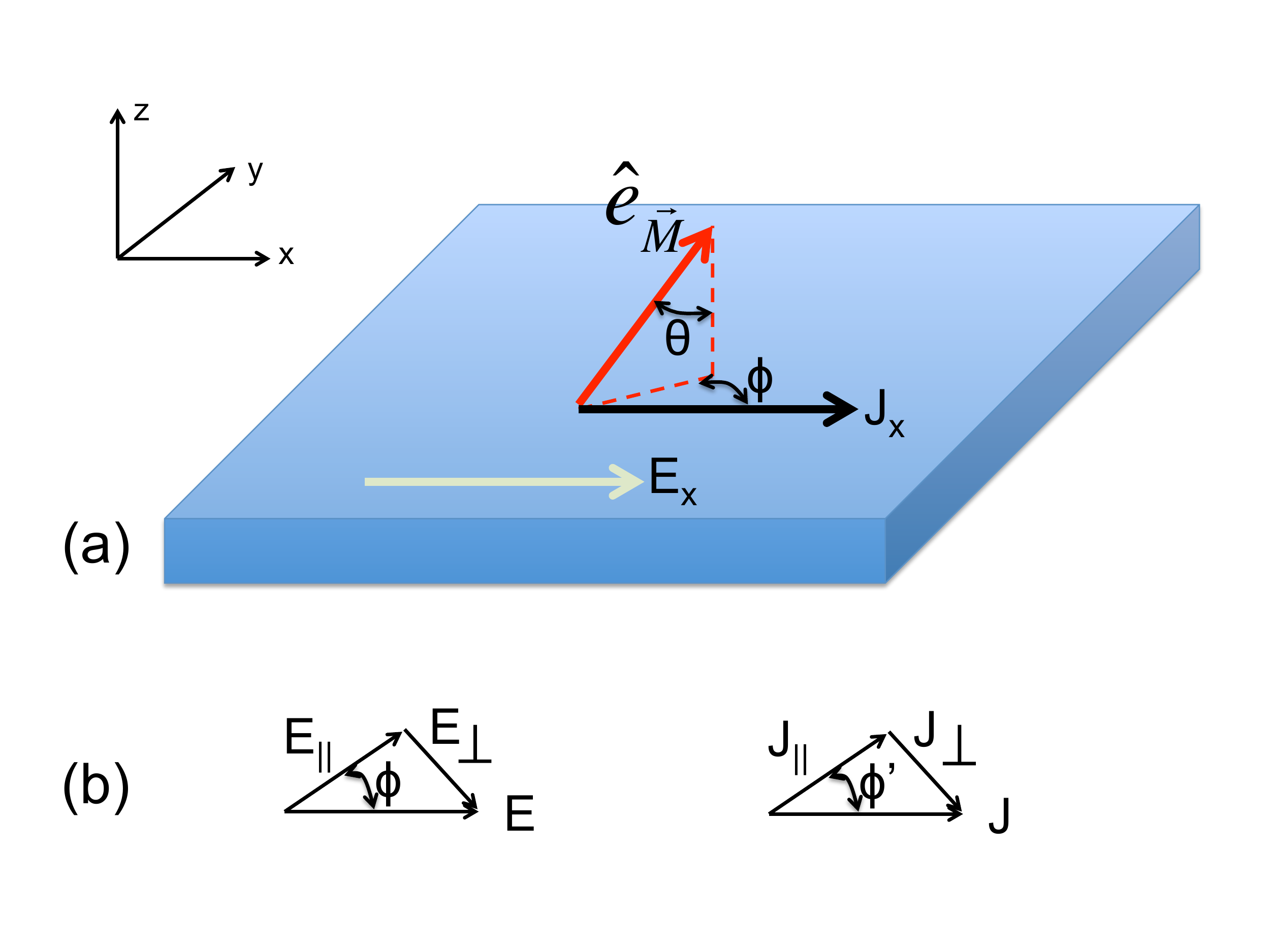}
\caption{(a) Geometry of the system considered: a magnetic moment rotated by a polar angle $\theta$ and azimutal angle 
$\phi$. The current density $J$ is related to the electric field $E$ via the resistivity. (b) Decomposition of the electric 
field and the current density parallel and perpendicular to the in-plane projection of the unit vector of the magnetic moment 
$\hat{e}_{\vec{\mathrm{M}}}$. }%
\label{Figure_appendix}%
\end{figure}

Here we derive phenomenologically the functional forms, which fit the 
computed longitudinal and transversal components of the residual 
resistivity tensor. The system of interest is an adatom with a tilted magnetic moment 
interacting with a gas of Rashba electrons. We assume a 2D current density flowing along the $x$-direction, $J_x$, that 
generates an electric field $\vec{E}$ (see Fig.\ref{Figure_appendix}(a)). 
Before analysing the general case of a tilted magnetic moment, let us 
recap what is expected when 
(i) the moment lies in-plane and (ii) the moment points out-of-plane. In case 
(i), we proceed as done by Thompson et al.~\cite{D.A.Thompson:1975} and consider the $x$-component of $\vec{E}$:
\begin{eqnarray}
E_x^{\mathrm(i)} &=& E_{\parallel} \cos\phi_{\vec{\mathrm{M}}} + E_{\perp} \sin\phi_{\vec{\mathrm{M}}}
\end{eqnarray}
where $E_{\parallel}$ and $E_{\perp}$ are the components of the electric field parallel and 
perpendicular to the projection of the unit vector of the magnetic moment, $\hat{e}_{\vec{\mathrm{M}}}$, 
lying in the $(xy)$ plane (the surface plane) 
as depicted in Fig.\ref{Figure_appendix}(b).
In terms of the current density and resistivity, the previous equation is then rewritten considering the 
parallel and perpendicular projection of the 2D current density on the direction of the magnetic moment:
\begin{eqnarray}
E_x^{\mathrm(i)} &=&  J_{\parallel} \rho_{\parallel} \cos\phi_{\vec{\mathrm{M}}} + 
J_{\perp} \rho_{\perp} \sin\phi_{\vec{\mathrm{M}}} 
\end{eqnarray}
as function of the azimutal angle $\phi_{\vec{\mathrm{M}}}$. Also, knowing that $J = J_{\parallel} \cos\phi_{\vec{\mathrm{M}}} = J_{\perp} \cos\phi_{\vec{\mathrm{M}}}$ (see Fig.\ref{Figure_appendix}(b)) leads to:
\begin{eqnarray}
E_x^{\mathrm(i)} &=&  J \large( \rho_{\parallel} \cos^2\phi_{\vec{\mathrm{M}}} + 
 \rho_{\perp} \sin^2\phi_{\vec{\mathrm{M}}} \large).
\end{eqnarray}

 Here, though, we give this expression in terms of 
the unit vector, $\hat{e}_{\vec{\mathrm{M}}}$, defining the orientation of the moment:
\begin{eqnarray}
E_x^{\mathrm(i)} &=&  J \large( \rho_{\parallel} (\hat{e}_{\vec{\mathrm{M}}} \cdot \hat{e}_x)^2 + 
 \rho_{\perp} (\hat{e}_{\vec{\mathrm{M}}} \cdot \hat{e}_y)^2 \large).
\end{eqnarray} 

Our proposal is that in the general case of a tilted magnetic moment, the previous two equations involving $\hat{e}_{\vec{\mathrm{M}}} \cdot \hat{e}_{x/y}$ holds. However, 
there is a missing contribution from the out-of-plane component of the magnetic moment. In the extreme case (ii), i.e. magnetic moment out-of-plane, we have:
\begin{eqnarray}
E_x^{\mathrm(ii)} &=& \rho^{\mathrm(ii)} J  
\label{Ex_i}
\end{eqnarray}
and a simple generalization leads to:
\begin{eqnarray}
E_x^{\mathrm(ii)} &=& \rho^{\mathrm(ii)} J  (\hat{e}_{\vec{\mathrm{M}}} \cdot \hat{e}_z)^2  
\end{eqnarray} 

As deduced from our numerical investigation, $\rho_{\parallel}=\rho^{\mathrm(ii)}$ (see FIG.~\ref{3D_res} (a)). 
This can be explained from  FIG.~\ref{Fermi_surface} (b) 
for $\vec{\mathrm{M}} \parallel z$ and FIG.~\ref{Fermi_surface} (c) for $\vec{\mathrm{M}} \parallel x$, where the 
allowed scattering processes are the same except 
for the interband scattering which flips the spin but does not change the direction of $\vec{k}$. 
The latter affects only the spin part of the response function, not the residual resistivity (charge part) that 
we compute. Therefore we get
the same residual resistivity for $\vec{\mathrm{M}} \parallel z$ and $\vec{\mathrm{M}} \parallel x$.

Now we can add up  both contributions (i) and (ii) and find:
\begin{eqnarray}
E_x &=&  J \large( \rho_{\parallel} \hat{e}_{\vec{\mathrm{M}}} \cdot (\hat{e}_x + \hat{e}_z)^2) +
 \rho_{\perp} (\hat{e}_{\vec{\mathrm{M}}} \cdot \hat{e}_y)^2 \large) ,
\end{eqnarray} 
which simplifies into:
\begin{eqnarray}
E_x &=&  J \large( \rho_{\parallel} + 
(\rho_{\perp}-\rho_{\parallel})\sin^2\phi_{\vec{\mathrm{M}}}\sin^2\theta_{\vec{\mathrm{M}}}\large).
\end{eqnarray} 

A similar approach can be used to derive the functional forms for the transversal part of the residual resistivity 
tensor. Here we address the $y$-component of $\vec{E}$ and again after starting from the form of 
Thompson et al.\cite{D.A.Thompson:1975} for the case (i):
\begin{eqnarray}
E_y^{\mathrm(i)} &=& E_{\parallel} \sin\phi_{\vec{\mathrm{M}}} - E_{\perp} \cos\phi_{\vec{\mathrm{M}}}, 
\end{eqnarray}
we end up with
\begin{eqnarray}
E_y^{\mathrm(i)} &=&  J \large( \rho_{\parallel} (\hat{e}_{\vec{\mathrm{M}}} \cdot \hat{e}_y)^2  - 
 \rho_{\perp} (\hat{e}_{\vec{\mathrm{M}}} \cdot \hat{e}_x)^2 \large). 
\end{eqnarray}
Since there is no transversal resistivity in the case (ii), the contribution $E_y^{\mathrm(ii)}$ vanishes 
and we find:
\begin{eqnarray}
E_y &=& J(\rho_{\perp}-\rho_{\parallel}) \cos\phi_{\vec{\mathrm{M}}} \sin\phi_{\vec{\mathrm{M}}} 
\sin^2\theta_{\vec{\mathrm{M}}}\large).
\end{eqnarray}

\end{document}